\documentclass[fleqn,usenatbib]{mnras}

\usepackage{newtxtext,newtxmath}

\usepackage[T1]{fontenc}

\DeclareRobustCommand{\VAN}[3]{#2}
\let\VANthebibliography\thebibliography
\def\thebibliography{\DeclareRobustCommand{\VAN}[3]{##3}\VANthebibliography}

\newcommand{\hi}{\textrm{H\textsc{i}}}
\newcommand{\secref}[1]{\hyperref[#1]{Section~\ref*{#1}}}
\newcommand{\appref}[1]{\hyperref[#1]{Appendix~\ref*{#1}}}
\newcommand{\txteq}[1]{\,{#1}\,}
\newcommand{\Nfg}{N_\text{fg}}
\newcommand{\hMpc}{\,h\,\text{Mpc}^{-1}}
\newcommand{\deltadiff}{\delta\hspace{-0.2mm}}

\usepackage{graphicx}	
\usepackage{enumitem}

\title[The foreground transfer function for \hi\ IM]
{The foreground transfer function for \hi\ intensity mapping signal reconstruction: MeerKLASS and precision cosmology applications}

\author[S. Cunnington et al.]{Steven Cunnington$^{1}$\thanks{E-mail: \href{mailto:steven.cunnington@manchester.ac.uk}{\texttt{steven.cunnington@manchester.ac.uk}}}\thanks{On behalf of the MeerKLASS Collaboration},
Laura Wolz$^{1}$,
Philip Bull$^{1,2}$,
Isabella P. Carucci$^{3,4}$,
Keith Grainge$^{1}$, 
Melis O. Irfan$^{2,5}$,
\newauthor
Yichao Li$^{6,2}$,
Alkistis Pourtsidou$^{7,8,2}$,
Mario G. Santos$^{2,9}$,
Marta Spinelli$^{10,2}$,
Jingying Wang$^{11,2}$
\\
$^1$Jodrell Bank Centre for Astrophysics, Department of Physics \& Astronomy, The University of Manchester, Manchester M13 9PL, UK\\
$^2$Department of Physics and Astronomy, University of the Western
Cape, Robert Sobukwe Road, Cape Town, 7535, South Africa\\
$^3$Dipartimento di Fisica, Universit\`a degli Studi di Torino, via P. Giuria 1, 10125, Torino, Italy\\
$^4$INFN – Istituto Nazionale di Fisica Nucleare, Sezione di Torino, via P. Giuria 1, 10125, Torino, Italy\\
$^{5}$Department of Physics and Astronomy, Queen Mary University of London, London, E1 4NS, UK\\
$^6$Department of Physics, College of Sciences, Northeastern University, Wenhua Road, Shenyang, 11089, China\\
$^7$Institute for Astronomy, The University of Edinburgh, Royal Observatory, Edinburgh EH9 3HJ, UK\\
$^{8}$Higgs Centre for Theoretical Physics, School of Physics and Astronomy, The University of Edinburgh, Edinburgh EH9 3FD, UK\\
$^9$South African Radio Astronomy Observatory (SARAO), 2 Fir Street, Cape Town, 7925, South Africa\\
$^{10}$Institute of Particle Physics \& Astrophysics, Department of Physics, ETH Zurich, Switzerland\\
$^{11}$Shanghai Astronomical Observatory, Chinese Academy of Sciences, 80 Nandan Road, Shanghai, 200030, China\\
}

\date{Accepted XXX. Received YYY; in original form ZZZ}

\pubyear{2023}

\begin{document}
\label{firstpage}
\pagerange{\pageref{firstpage}--\pageref{lastpage}}
\maketitle

\begin{abstract}
Blind cleaning methods are currently the preferred strategy for handling foreground contamination in single-dish \hi\ intensity mapping surveys. Despite the increasing sophistication of blind techniques, some signal loss will be inevitable across all scales. Constructing a corrective transfer function using mock signal injection into the contaminated data has been a practice relied on for \hi\ intensity mapping experiments. However, assessing whether this approach is viable for future intensity mapping surveys where precision cosmology is the aim, remains unexplored. In this work, using simulations, we validate for the first time the use of a foreground transfer function to reconstruct power spectra of foreground-cleaned low-redshift intensity maps and look to expose any limitations. We reveal that even when aggressive foreground cleaning is required, which causes ${>}\,50\%$ negative bias on the largest scales, the power spectrum can be reconstructed using a transfer function to within sub-percent accuracy. We specifically outline the recipe for constructing an unbiased transfer function, highlighting the pitfalls if one deviates from this recipe, and also correctly identify how a transfer function should be applied in an auto-correlation power spectrum. We validate a method that utilises the transfer function variance for error estimation in foreground-cleaned power spectra. Finally, we demonstrate how incorrect fiducial parameter assumptions (up to ${\pm}100\%$ bias) in the generation of mocks, used in the construction of the transfer function, do not significantly bias signal reconstruction or parameter inference (inducing ${<}\,5\%$ bias in recovered values). 
\end{abstract}




\section{Introduction}

Probing fluctuations in the Universe's density field is an excellent tool for furthering precision cosmology. A number of large sky surveys have been commissioned with this aim and have contributed towards constraining parameters in the standard cosmological model \citep{eBOSS:2020yzd,Heymans:2020gsg,DES:2021wwk}. Despite cosmic microwave background (CMB) experiments leading the way with constraints \citep{Aghanim:2018eyx}, it is expected that large-scale structure maps will soon be the leading resource given the three-dimensional information they provide \citep{Slosar:2008hx,Giannantonio:2011ya,Alonso:2015uua}. Increases in survey volume will allow fluctuations across the largest scales to be probed, improving constraints. It is within the relatively unexplored \textit{ultra}-large scales where novel tests of general relativity will be possible and where there will be increased sensitivity to new physics such as non-Gaussian fluctuations in the Universe's primordial density field  \citep{Camera:2013kpa,Camera:2014sba,Alvarez:2014vva,Fonseca:2015laa,Baker:2015bva,Bull:2015lja,Cunnington:2022ryj}. Ultra-large scales are also highly linear, avoiding the complex modelling challenges facing surveys attempting to exploit non-linear regimes \citep{DAmico:2019fhj,Euclid:2020tff, Pourtsidou:2022gsb}.

An efficient method for surveying large volumes is using radio telescopes to map the redshifted 21cm emission from neutral hydrogen (\hi). The \hi, which mostly resides in galaxies in the post-reionization Universe, traces the underlying dark matter, thus allowing the Universe's large-scale structure to be probed. By rapidly scanning the sky and recording all radiation as unresolved diffuse emission, the faint 21cm signals are integrated allowing for a comprehensive survey of \hi\ density in 3D to be obtained. This process is known as \hi\ intensity mapping \citep{Bharadwaj:2000av,Battye:2004re,Wyithe:2007rq,Chang:2007xk}. 

The \hi\ power spectrum has been recently detected on small Mpc scales \citep{Paul:2023yrr} with intensity maps from the 64-dish MeerKAT array, a pathfinder telescope for the Square Kilometre Array Observatory (SKAO) \citep{Bacon:2018dui}. This detection used MeerKAT as an interferometer which has higher sensitivity on small scales. However, within the next few years the MeerKAT Large Area Synoptic Survey (MeerKLASS) plans to conduct a wide (${\gtrsim}\,4{,}000\,\text{deg}^2$) \hi\ intensity mapping survey, potentially spanning $0.4\txteq{<}z\txteq{<}1.45$ in redshift if performed using the UHF band \citep{MeerKLASS:2017vgf}.
Since the MeerKAT interferometer does not have sufficiently short baselines to achieve such a field of view, the observations will be gathered using the single-dish data from each element of the array. This auto-correlation mode of observation, often referred to as \textit{single-dish mode}, is also planned for the full SKAO in order to probe large-scale cosmology, which has been identified as a top priority science objective \citep{Weltman:2018zrl}. In the pre-SKA era however, MeerKAT will pursue low-redshift \hi\ intensity mapping and has already demonstrated calibration and map-making from single-dish mode observations with a small pilot survey \citep{Wang:2020lkn}. This same pilot survey was also used to achieve the first single-dish mode cosmological detection with a multi-dish array in cross-correlation with an overlapping galaxy survey \citep{Cunnington:2022uzo}.

Since \hi\ intensity mapping records all emission in the frequency range of the instrument, the major challenge is removing any signals which are not cosmological \hi. This can include radio frequency interference (RFI) and astrophysical foregrounds, both of which can dominate by orders of magnitude over the weak \hi\ signal\footnote{Additional contaminants come from atmospheric emission and ground pickup which can be approximately modelled as constant over time when a constant elevation scanning strategy is adopted.}. In principle, RFI should be time-varying and can be flagged when it enters the observations. However, foregrounds will consistently enter the observations due to their fixed sky coordinates, therefore a process for separating them from the \hi\ is required. The dominant sources producing foregrounds in the low redshift \hi\ frequency range (${\sim}\,300\txteq{<}\nu\txteq{<}1420$\,\text{MHz}) are synchrotron and free-free radiation from within our own galaxy, along with extra-galactic concentrated emission from strong point sources such as active galactic nuclei \citep{Oh:2003jy,Santos:2004ju, Alonso:2014sna}.

To date, \textit{blind} foreground cleaning techniques have been the only approach that has led to cross-correlation detections of a cosmological power spectrum in single-dish intensity mapping \citep{Masui:2012zc,Wolz:2015lwa,Anderson:2017ert,Wolz:2021ofa,Cunnington:2022uzo}. Blind techniques exploit the robust assumption that foregrounds are a dominant and correlated contribution to the observations and can be statistically reduced into a few components which are removed. This requires little prior knowledge of the foregrounds which is a huge advantage since it is challenging to obtain a detailed understanding of the foreground's precise amplitude through frequency, or how they respond to instrumental systematics. Blind foreground removal performed at the map level has proven to be the most successful approach and these have been validated and refined in simulations \citep{Wolz:2013wna,Alonso:2014dhk,Carucci:2020enz,Cunnington:2020njn,Spinelli:2021emp}. Interferometric intensity mapping can to some extent adopt a foreground avoidance strategy, which assumes they are isolated in a foreground wedge region in ($k_\perp,k_\parallel$)-Fourier space \citep{Liu:2014yxa,Paul:2023yrr}. The foreground avoidance technique has the advantage of being immune to signal loss from foreground cleaning. However, recent studies have shown some component separation improves foreground mitigation relative to foreground avoidance alone in interferometric surveys \citep{Chen:2022viv}. Hence blind foreground removal is likely to be an adopted technique beyond single-dish mode experiments. 

Whilst blind foreground cleaning algorithms themselves have been well studied, the precise effects they cause on the underlying \hi\ field have not been to the same extent. Broadly speaking there are two unfavourable consequences from blind foreground cleaning, and both will occur simultaneously to some extent. The first is \textit{foreground residuals}, i.e. foreground contamination not removed from the data resulting from under-cleaning. The second is \textit{signal loss} i.e. the reduction in the \hi\ power spectrum amplitude resulting from the foreground clean. It is this second issue that is the main focus of this paper. Whilst foreground residuals are of course important, their influence on the data is similar to RFI and instrumental noise, that is they cause an additive bias to the estimated \hi\ power spectrum. However, for foregrounds, it is expected that their residuals should be reducible to sub-dominant levels relative to the \hi\ \citep{Cunnington:2020njn}. Furthermore, in cross-correlation with a foreground-free tracer such as a galaxy survey, any additive bias from foreground residuals will be absent and the only impact will be on the error budget.

For signal loss, it has been shown that even in ideal simulations, some loss is always inevitable across all scales when blind foreground removal methods are applied, and this is not mitigated in cross-correlations \citep{Cunnington:2020njn}. Ignoring or incorrectly estimating signal loss, unsurprisingly, leads to a biased recovery of the \hi\ power spectrum. Thus signal loss is a crucial concept to understand exhaustively for precision cosmology to be possible with \hi\ intensity mapping. The necessary process of signal reconstruction i.e. correcting for the signal loss, is where there is little dedicated study. A foreground transfer function $\mathcal{T}$ can be simply defined as the object which delivers a reconstructed signal power spectrum that is unbiased to the underlying truth i.e. $\langle P_\text{rec}(k)\rangle\txteq{=}P_\text{true}(k)$ where $\langle P_\text{rec}(k)\rangle\txteq{\equiv}P_\text{clean}(k)/\mathcal{T}(k)$. Previous intensity mapping detections \citep{Masui:2012zc,Anderson:2017ert,Wolz:2021ofa,Cunnington:2022uzo} have all relied on a process of mock signal injection to estimate the foreground transfer function. By subjecting the injected mocks to the same foreground cleaning process as the observations, we can use the drop in the measured mock power spectra to estimate the transfer function. This method was first extensively analysed in \citet{Switzer:2015ria} in the context of the Green Bank Telescope (GBT) \hi\ intensity maps \citep{Masui:2012zc,Switzer:2013ewa}. There have been similar methods of signal injection to correct for signal loss implemented for epoch of reionisation experiments where past analyses underestimated signal loss leading to biased results \citep{Cheng:2018osq}, highlighting the importance of correctly understanding this issue. To date, there has been no dedicated simulations-based investigation into the reliability of the transfer function for low-redshift \hi\ intensity maps blindly cleaned at the map level, despite its clear importance.

In this work, we use various \hi\ intensity mapping simulations to validate the reliability of the transfer function for signal reconstruction in foreground-cleaned maps. We explore how signal loss arises in a foreground clean and illustrate the subtleties of this both analytically and empirically in simulation tests, demonstrating how a transfer function can be correctly estimated to account for all these subtleties. We focus exclusively on Principal Component Analysis (PCA)-based foreground cleaning but much of the formalism and results presented will be transferable to other foreground cleaning techniques. Furthermore, whilst our focus is on single-dish intensity mapping, the conclusions will also be applicable to interferometers. We demonstrate how a foreground transfer function is a reliable tool for small pilot surveys, validating it on simulations constructed using empirical MeerKAT 2019 observations aiming to realistically emulate current MeerKAT intensity maps. Lastly, we look to the future and pursue to what extent the transfer function can be relied on for conducting precision cosmology with intensity mapping where sub-percent accuracy on parameter estimation is the aim.

This paper is structured as follows; in \secref{sec:signallossformalism} we present an overview of the formalism for a blind PCA-based foreground clean, explicitly highlighting where signal loss arises. \secref{sec:TF} presents how one should construct a foreground transfer function to correct for signal loss. In \secref{sec:MK} we test the transfer function on a simulation of a MeerKAT-like intensity mapping pilot survey, validating the transfer function in this low signal-to-noise regime. \secref{sec:cosmo} focuses on how suited the transfer function is for the purposes of precision cosmology, showcasing the robustness of the transfer function even where the fiducial cosmology assumed for its construction disagrees with the truth in the observations. Finally, we conclude in \secref{sec:Conclusion}.

\section{Signal loss from foreground cleaning}\label{sec:signallossformalism}

We begin with a pedagogical introduction to the formalism describing blind foreground cleaning and with the aid of simulations demonstrate some key concepts of signal loss induced by the foreground clean. For consistency, we largely follow the notation in \citet{Switzer:2015ria}. Whilst we focus on a PCA-based method, the formalism we present is in principle transferable to more sophisticated blind foreground removal techniques when they are used as linear filters \cite[e.g.][]{4337755,Chapman:2012yj,Alonso:2014dhk,Carucci:2020enz,Cunnington:2020njn,Irfan:2021bci,Spinelli:2021emp}. We \textcolor{black}{use} a set of simulations with separable \hi\ signal and foreground contributions, allowing us to provide examples of the claims made in certain derivations. We begin by using some generic simulations which are similar to that used in \citet{Cunnington:2020njn}. The exact details of these simulations are outlined in \appref{sec:MD1GPC} but we include a short summary of points below.

\begin{itemize}[leftmargin=*]
    \item The $1\,(\text{Gpc}/h)^3$ \textsc{MultiDark} \citep{Klypin:2014kpa} $N$-body semi-analytical simulation with approximate cold gas masses is used for the single realisation of the underlying \textit{true} \hi\ signal at a snapshot redshift of $z\txteq{=}0.39$, gridded into $n_\text{x},n_\text{y},n_\text{z}\txteq{=}256,256,256$ voxels. We include redshift-space distortions (RSD) to provide the \hi\ signal with an anisotropic signature. A frequency range of $900\txteq{<}\nu\txteq{<}1156\,\text{MHz}$ with resolution $\deltadiff\nu\txteq{=}1\,\text{MHz}$ is assumed which is consistent with the snapshot redshift and is reasonably consistent with MeerKAT L-band intensity mapping observations.

    \item We simulate galactic synchrotron, galactic free-free, and bright point source emission at these frequencies to provide a foreground sky. We use the Planck Legacy Archive\footnote{\href{http://pla.esac.esa.int/pla}{pla.esac.esa.int/pla}} FFP10 simulations for the synchrotron and free-free emission. The point sources catalogue is produced following the same approach as in \citet{batps}.
    
    \item We cut a patch of sky consistent with the $1\,(\text{Gpc}/h)^2$ \hi\ survey size and chose this to be centered on the Stripe~82 region of the sky, where a real survey could be targeted. The foreground component with $n_\theta\txteq{=}n_\text{x}\txteq{\times}n_\text{y}$ angular pixels and $n_\nu\txteq{\equiv}n_\text{z}$ frequency channels is added onto the \hi\ signal simulation.

    \item To increase the complexity of the foreground clean, we simulate instrumental polarisation leakage \citep{Carucci:2020enz,Cunnington:2020njn} which disrupts the smooth frequency spectra of the foreground simulations, requiring a clean which is more aggressive and consistent with real data. For this we used the \texttt{CRIME}\footnote{\href{http://intensitymapping.physics.ox.ac.uk/CRIME.html}{intensitymapping.physics.ox.ac.uk/CRIME.html}} software \citep{Alonso:2014sna}. This is used by default and we highlight any cases where this has not been used.

    \item By default we add no further instrumental effects, but in some cases we introduce instrumental noise and smoothing perpendicular to the line-of-sight to emulate the telescope beam. For the noise, we assume isotropic Gaussian white noise with $\sigma_\text{n}\txteq{=}1\,\text{mK}$, approximately corresponding to $30\,\text{hrs}$ of observation time on a MeerKAT-like survey of ${\sim}\,3{,}000\,\text{deg}^2$ sky (see \autoref{eq:noise} and \ref{eq:timeperpoint} for more details). This noise dominates over the \hi\ signal which has an rms of $\sigma_\hi\txteq{\sim}0.14\,\text{mK}$. The beam we approximate as a Gaussian with comoving transverse length scale $R_\text{beam}\txteq{=}10\,h^{-1}\text{Mpc}$ (see \autoref{eq:GaussianSpatialBeam} for a definition). We clearly indicate cases where noise or a beam has been added. We discuss the limitations of these approximations and also explore some more realistic systematics in \secref{sec:MK} based on real MeerKAT pathfinder data, which we introduce there.
    
\end{itemize}

Throughout we will refer to these as the MD1GPC simulations. Later in the paper, we use some more specific simulations to explore different scenarios which we will introduce then, but for the majority of our results, we use the MD1GPC by default unless otherwise clearly stated.\\

\noindent To begin a PCA-based clean of the \hi\ + foregrounds combination, we first calculate the $\nu,\nu'$ covariance of the foreground contaminated data $\textbf{\textsf{X}}_\text{obs}$, where the data matrices $\textbf{\textsf{X}}$ have dimensions $[n_\nu,\,n_\theta]$. The covariance is estimated by $\textbf{\textsf{C}}\txteq{=}(n_{\theta}\txteq{-}1)^{-1} \textbf{\textsf{X}}_\text{obs}^\text{T}\textbf{\textsf{X}}_\text{obs}\txteq{=}\textbf{\textsf{U}}\boldsymbol{\Lambda}\textbf{\textsf{U}}^\text{T}$, where the last equality is the eigen-decomposition (or diagonalisation) of the covariance matrix, with $\textbf{\textsf{U}}$ representing a matrix with the $n_\nu$ spectral eigenvectors $\textbf{\textsf{U}}_i$ and $\boldsymbol{\Lambda}$ is the diagonal matrix of eigenvalues. Neglecting noise contributions,\footnote{This is mainly done for brevity and incorporating noise into this formalism is not overly complicated. It acts as additional perturbations to the pure foreground modes in a similar way to the \hi\ signal, as we later discuss.} i.e. $\textbf{\textsf{X}}_\text{obs}\txteq{=}\textbf{\textsf{X}}_\text{f+s}\txteq{\equiv}\textbf{\textsf{X}}_\text{f}\txteq{+}\textbf{\textsf{X}}_\text{s}$, we can write this as
\begin{equation}
    \textbf{\textsf{C}} = (n_{\theta}\,{-}\,1)^{-1} \left(\textbf{\textsf{X}}_\text{f}+\textbf{\textsf{X}}_\text{s}\right)^\text{T}\left(\textbf{\textsf{X}}_\text{f}+\textbf{\textsf{X}}_\text{s}\right)\,,
\end{equation}
which expands to
\begin{equation}\label{eq:PerturbedCovariance}
\begin{aligned}
    \textbf{\textsf{C}}_{\mathrm{f}+\mathrm{s}} & = (n_{\theta}\,{-}\,1)^{-1}\left(\textbf{\textsf{X}}_{\mathrm{f}} \textbf{\textsf{X}}_{\mathrm{f}}^\text{T}+\textbf{\textsf{X}}_{\mathrm{f}} \textbf{\textsf{X}}_{\mathrm{s}}^\text{T}+\textbf{\textsf{X}}_{\mathrm{s}} \textbf{\textsf{X}}_{\mathrm{f}}^\text{T}+\textbf{\textsf{X}}_{\mathrm{s}} \textbf{\textsf{X}}_{\mathrm{s}}^\text{T}\right) \\
    &=\textbf{\textsf{C}}_{\mathrm{f}}+\textbf{\textsf{C}}_{\Delta}\,,
\end{aligned}
\end{equation}
where $\textbf{\textsf{C}}_{\Delta}\,{=}\,(n_{\theta}\,{-}\,1)^{-1}(\textbf{\textsf{X}}_{\mathrm{f}} \textbf{\textsf{X}}_{\mathrm{s}}^\text{T}+\textbf{\textsf{X}}_{\mathrm{s}} \textbf{\textsf{X}}_{\mathrm{f}}^\text{T}+\textbf{\textsf{X}}_{\mathrm{s}} \textbf{\textsf{X}}_{\mathrm{s}}^\text{T})$ are residual contributions to the estimate of the foreground covariance. The estimate of the foregrounds, which is to be removed from the observations, is then given by
\begin{equation}\label{eq:FGest}
    \hat{\textbf{\textsf{X}}}_\text{f} = \textbf{\textsf{U}}\textbf{\textsf{S}}\textbf{\textsf{U}}^\text{T}\textbf{\textsf{X}}_\text{obs}\,, 
\end{equation}
where, following \citet{Switzer:2015ria}, we introduce the selection matrix $\textbf{\textsf{S}}$ which is zero everywhere except for the first $\Nfg$ elements along the diagonal which are set equal to one, assigning the number of contaminated modes projected out from each line of sight.

\begin{figure}
    \centering
    \includegraphics[width=1\linewidth]{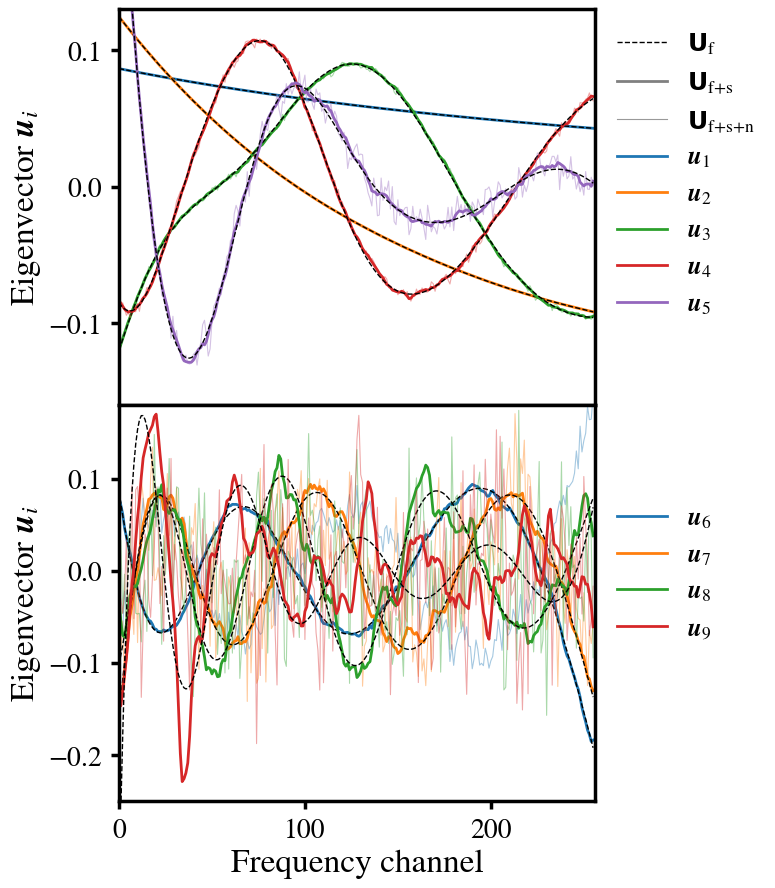}
    \caption{The first 9 eigenvectors (first 5 in the top panel and next four in the bottom) from the PCA on the MD1GPC foreground contaminated simulations. The dashed-black lines show purely foreground modes taken from the PCA on foreground-only sims. The solid colour lines show the eigenvectors perturbed by the inclusion of the \hi\ signal in the data. These perturbations are the origins of signal loss in foreground cleaning. The thin-faint solid color lines show eigenvectors perturbed by the inclusion of the \hi\ signal + noise.}
    \label{fig:PerturbedEigenvecs}
\end{figure}

Whilst we expect the foregrounds to be orders of magnitude larger than the \hi\ signal, and $\textbf{\textsf{C}}_{\mathrm{f}}$ to be the dominant term in \autoref{eq:PerturbedCovariance}, it is important for the additional perturbations from the signal through $\textbf{\textsf{C}}_{\Delta}$ to be considered. Due to this mix of foregrounds and signal, the eigenvectors we obtain are 
\begin{equation}\label{eq:PerturbedEigenvectors}
    \textbf{\textsf{U}}\equiv\textbf{\textsf{U}}_{\mathrm{f}+\mathrm{s}}=\textbf{\textsf{U}}_{\mathrm{f}}+\boldsymbol{\Delta}\,,
\end{equation}
where $\textbf{\textsf{U}}_{\mathrm{f}}$ are pure foreground modes and $\boldsymbol{\Delta}$ are the perturbed contributions caused by the signal. We show the difference between the unperturbed ($\textbf{\textsf{U}}_\text{f}$) and perturbed ($\textbf{\textsf{U}}_\text{f+s}$) eigenvectors in \autoref{fig:PerturbedEigenvecs}. This shows the first 9 most dominant eigenvectors from the MD1GPC simulations. For the unperturbed, pure foreground eigenvectors (dashed black lines), the modes are smooth in frequency, only showing longer wavelength oscillations caused by the polarisation leakage. However, for the eigenvectors estimated from the foreground and signal mix (solid coloured lines), perturbations to the modes caused by the signal start to arise. These perturbations are more severe the higher the order of the eigenvector. This is because the eigenvectors become increasingly less dominant and are more easily perturbed by the presence of the signal whose contributions remain fairly consistent through all the eigenvectors due to its high-rank properties.

\autoref{fig:PerturbedEigenvecs} begins to demonstrate how signal loss can enter in a foreground clean. By using the eigenvectors $\textbf{\textsf{U}}_\text{f{+}s}$ as the basis functions which are projected out in the PCA clean, this will project out modes that have some \hi\ structure shown by the perturbations on the lower modes. It is tempting to try and address this issue at this early stage and use a low-pass filter or smooth the perturbed modes to correct the perturbations from the signal. However, we briefly experimented with various smoothing routines with this aim and in all cases, the results were made worse. Since the aim of this work is not to enhance foreground cleaning efficiency, but instead to ensure we can control signal loss, we defer any investigations into cleaning optimisation to future work. 

We also show in \autoref{fig:PerturbedEigenvecs} the impact on the eigenmodes by including the dominant instrumental noise ($\textbf{\textsf{U}}_\text{f+s+n}$ shown by faint colored lines). These create much larger perturbations to the pure foreground modes, hence large noise can impact foreground cleaning. This is an important issue for early pilot surveys where observation time is low since in these cases, the noise will dominate over the \hi\ and will be the main source of perturbations to the eigenvectors. We will discuss this in more detail later and demonstrate how intensity maps with a high level of noise, and other additive systematics like residual RFI, can still undergo signal reconstruction using a transfer function. For the remainder of this section, we omit the instrumental noise for simplicity.

The perturbations in \autoref{fig:PerturbedEigenvecs} are dependent on the ratio between the foreground amplitude and the other components e.g. the \hi. Whilst the \hi\ signal amplitude will be consistently uniform due to the cosmological principle, the foreground emission can vary with the choice of sky patch, e.g. being orders of magnitude higher near the galactic plane relative to the South Celestial Pole. This was explored in \citet{Cunnington:2020njn} where different foreground regions were tested. For the remainder of this work, however, we will stick to one region as most of the conclusions we draw are generic regardless of how strong the foregrounds are, within a physically reasonable range.

\subsection{Toy model foreground cleaning}

Here we investigate some idealised foreground cleaning scenarios to demonstrate the nature of signal loss in blind foreground cleaning. We begin with the most ideal toy case scenario where we project out pure foreground modes from pure foreground-only data and subtract this from the observed combination $\textbf{\textsf{X}}_\text{f+s}$. Of course, if we could access perfect foreground-only data $\textbf{\textsf{X}}_\text{f}$ this could be simply subtracted from the observed foreground and signal mix, so there would be no need for any mode projection cleaning process. However, we proceed with this example since it provides valuable insight from which we can add complication. This first toy-case is given by
\begin{equation}\label{eq:IdealX_clean}
    \textbf{\textsf{X}}_\text{toy:clean1} = \textbf{\textsf{X}}_\text{f+s}-\textbf{\textsf{U}}_\text{f}\textbf{\textsf{S}}\textbf{\textsf{U}}_\text{f}^\text{T}\textbf{\textsf{X}}_\text{f}\,,
\end{equation}
In this ideal case, since we are projecting out perfect foreground modes from pure foreground data, the optimal selection matrix $\textbf{\textsf{S}}$ would have the diagonals filled with ones ($\Nfg\txteq{=}n_\nu$) i.e. the identity matrix, to remove all foreground without the consequence of signal loss. In reality, this is not possible and a balance is sought between projecting out enough modes to remove foregrounds but not so many that large signal loss is sustained.

\begin{figure*}
        \centering
    \includegraphics[width=\linewidth]{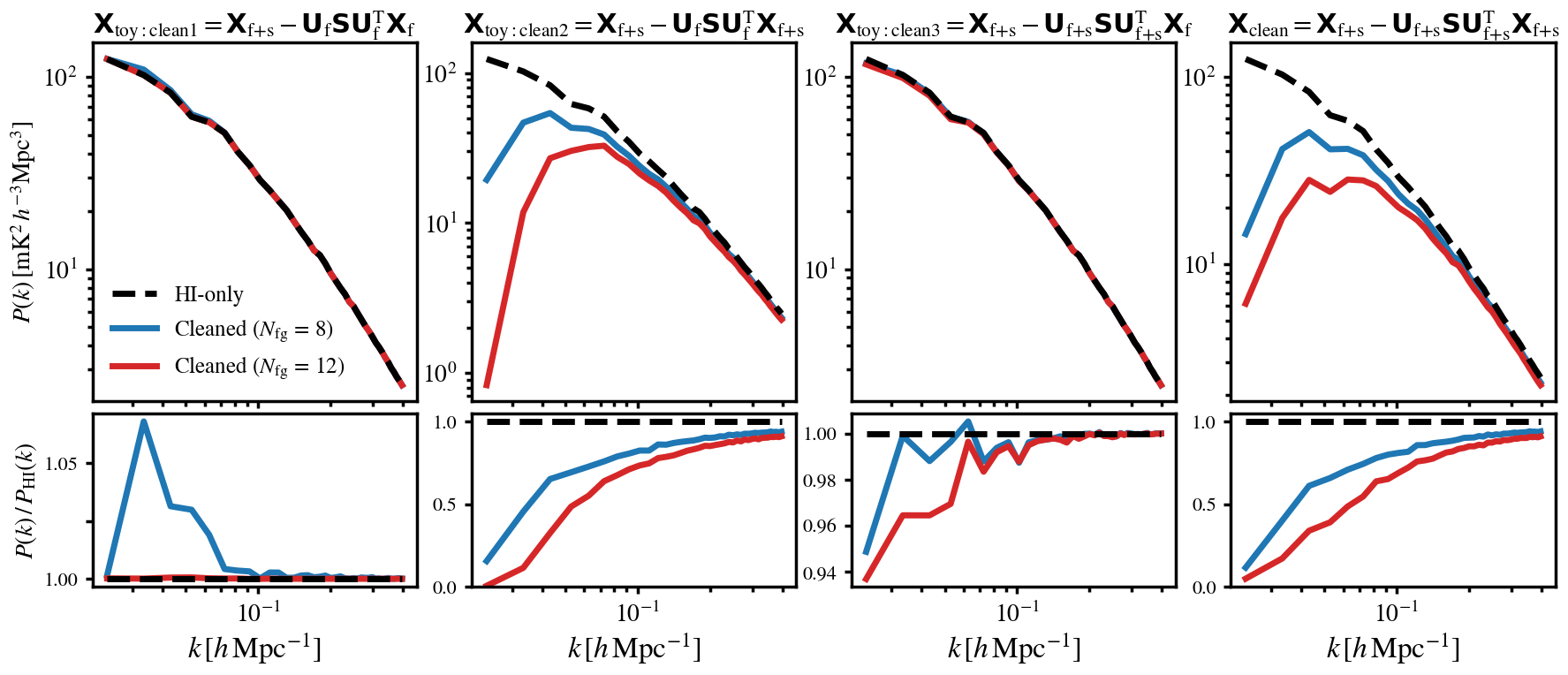}
    \caption{Auto power spectra for different foreground cleaning cases in comparison with the true \hi-only simulations (black-dashed line). The first three panels represent idealised scenarios. \textit{Far-left} shows projecting pure foreground modes ($\textbf{\textsf{U}}_\text{f}$) out from pure foreground data ($\textbf{\textsf{X}}_\text{f}$). \textit{Second panel} shows projecting pure foreground modes out from foreground data mixed with signal ($\textbf{\textsf{X}}_\text{f+s}$). \textit{Third panel} shows projecting modes perturbed by the signal ($\textbf{\textsf{U}}_\text{f+s}$) out from pure foreground data. \textit{Far-right} shows the realistic scenario where modes are perturbed by the signal and these are then projected out from foreground data mixed with signal.}
    \label{fig:Xclean_comb}
\end{figure*}

In \autoref{fig:Xclean_comb} we show the measured power spectrum for the idealised toy clean in \autoref{eq:IdealX_clean} (far-left panel). We also show other cases of foreground clean in the other panels which we will discuss shortly. Details on the power spectrum estimation formalism used throughout the paper are presented in \appref{sec:PowerSpec}. We show the power spectra in comparison with the original \hi-only (foreground-free) data, which we are aiming to agree with. In all cases we show two cleaned examples with 8 and 12 modes projected out, i.e. the first $\Nfg\txteq{=}8$ and $\Nfg\txteq{=}12$ elements in the diagonal of $\textbf{\textsf{S}}$ are set to 1. This is why we do not reach perfect agreement in the ideal top-left panel, because we are not projecting out all pure foreground modes, leaving some foreground residual in the remaining modes. Thus, there is slight disagreement with the \hi-only, albeit at a sub-percent level for the $\Nfg\txteq{=}12$ case. The values of $\Nfg\txteq{=}8,\,12$ are chosen to sufficiently suppress the polarised foregrounds in the increasingly more realistic cases shown by the other panels which we discuss next.

In reality, the situation in \autoref{eq:IdealX_clean} is not possible, because we can not project out modes from foreground-only data $\textbf{\textsf{X}}_\text{f}$ because the observed data we have, $\textbf{\textsf{X}}_\text{f+s}$, is inherently mixed with signal. Signal loss begins to manifest in the case where we project out foreground modes from the observed data mix
\begin{equation}\label{eq:Xclean2}
    \textbf{\textsf{X}}_\text{toy:clean2} = \textbf{\textsf{X}}_\text{f+s} - \textbf{\textsf{U}}_\text{f}\textbf{\textsf{S}}\textbf{\textsf{U}}_\text{f}^\text{T}\textbf{\textsf{X}}_\text{f+s}\,. 
\end{equation}
This is still an idealised scenario since we are projecting out \textit{pure} foreground modes from the data. In reality, a further complication arises since the modes identified in the PCA will be perturbed by the presence of signal i.e. $\textbf{\textsf{U}}_\text{f}\txteq{\rightarrow}\textbf{\textsf{U}}_\text{f+s}$, which we will discuss shortly. Comparing the first panel with the second, where the difference is that in the former case, pure foreground modes are now being projected out from the mix of foreground and signal (\autoref{eq:Xclean2}), evidence of signal loss in the cleaned power spectrum begins to show. The signal loss is clearly dominating over any foreground residuals remaining in the data from only projecting out a finite number of modes. In other words, the small ${\sim}\,5\%$ additive bias in the far-left panel caused by foreground residuals is not seen in the second panel, due to a more dominant impact from signal loss. Of course if a smaller number for $\Nfg$ were chosen, foreground residuals would cause more of a problem.

The second panel of \autoref{fig:Xclean_comb} confirms that signal loss begins to manifest when modes (even purely foreground ones) are projected out of the data, which is a combination of foregrounds and signal. The reason for this is because signal will unavoidably have degeneracies with some foreground structure. Thus when a set of foreground functions are projected out of the data, some signal will leak into this subtraction, mainly large-scale (small-$k$) modes since these are most degenerate with the foreground structure which is highly correlated through frequency.

The third panel of \autoref{fig:Xclean_comb} shows a final idealised toy case where we are only projecting out modes from the pure foreground map, but the modes are now perturbed by the presence of signal, $\textbf{\textsf{U}}_\text{f+s}$ (see \autoref{eq:PerturbedEigenvectors}). This is something we have to deal with in reality where we can not form a perfect estimation for the foreground-only eigenmodes $\textbf{\textsf{U}}_\text{f}$, from the true observed data where foreground and signal are mixed. In this case, the eigenvectors themselves are perturbed and it is these perturbed modes that we project along the data;
\begin{equation}\label{eq:Xclean3}
    \textbf{\textsf{X}}_\text{toy:clean3} = \textbf{\textsf{X}}_\text{f+s} - \textbf{\textsf{U}}_\text{f+s}\textbf{\textsf{S}}\textbf{\textsf{U}}_\text{f+s}^\text{T}\textbf{\textsf{X}}_\text{f}\,.  
\end{equation}
This provides an interesting result with signal loss again appearing to be the more dominant effect with little evidence of additive bias from foreground residuals. However we are only projecting out modes from pure foreground data, so it seems counter-intuitive that there is signal loss. As we will explicitly show in the following section, this is caused by the perturbation to the modes from the presence of signal ($\textbf{\textsf{U}}_\text{f+s}$), which creates a complicated mix of subtracted terms that can have signal correlating and anti-correlating contributions, as identified in \citet{Switzer:2015ria}.

\subsection{The origins and subtleties of signal loss}\label{sec:OriginsSubtletiesSignalLoss}

Despite seeing signal loss in the second and third panels of \autoref{fig:Xclean_comb}, both cases still represent unrealistic scenarios. In reality, we see a combination of both where the presence of signal perturbs the eigenmodes ($\textbf{\textsf{U}}_\text{f+s}$) as well as complicating the clean since information is projected out from data which contain not just foregrounds, but signal too ($\textbf{\textsf{U}}_\text{f+s}\textbf{\textsf{S}}\textbf{\textsf{U}}_\text{f+s}^\text{T}\textbf{\textsf{X}}_\text{f+s}$). Thus, the true resulting cleaned data is given by
\begin{equation}\label{eq:Xclean4}
    \textbf{\textsf{X}}_\text{clean} = \textbf{\textsf{X}}_\text{f+s} - \textbf{\textsf{U}}_\text{f+s}\textbf{\textsf{S}}\textbf{\textsf{U}}_\text{f+s}^\text{T}\textbf{\textsf{X}}_\text{f+s}\,. 
\end{equation}
The result from this foreground clean is shown in the final far-right panel of \autoref{fig:Xclean_comb}. Results appear similar to $\textbf{\textsf{X}}_\text{toy:clean2}$ but some differences can be seen on large scales caused by the increased complexity of having perturbed eigenmodes ($\textbf{\textsf{U}}_\text{f+s}$). 

\begin{figure*}
    \centering
    \includegraphics[width=0.92\textwidth]{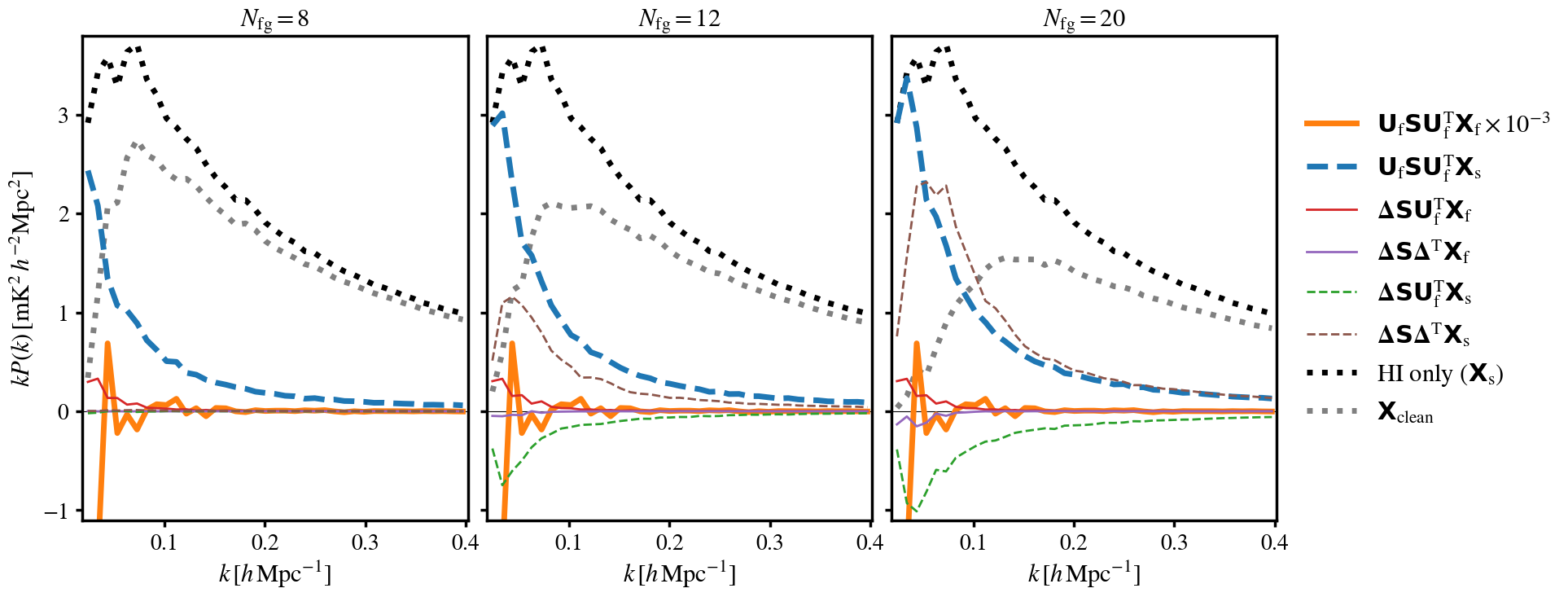}
    \caption{Contributions to signal loss from the decomposed terms in \autoref{eq:Xclean}. Each power spectra shows the cross-correlation between the original-\hi\ and a subtracted term in the foreground clean. The different panels represent different numbers of PCA modes removed ($\Nfg$). Solid lines indicate residual foreground contributions whereas dashed lines indicate \hi\ signal contributions. Thin lines indicate perturbed contributions from $\mathbf{\Delta}$ caused by the presence of signal in the eigenmode estimation. For reference, we also include the pure \hi-power (black-dotted) along with the total cleaned result $\textbf{\textsf{X}}_\text{clean}$ (grey-dotted).}
    \label{fig:DecomposedTerms}
\end{figure*}

To understand the complexity of foreground cleaning, we expand the above \autoref{eq:Xclean4} into all its terms, giving
\begin{equation}
\begin{aligned}
    \textbf{\textsf{X}}_{\text {clean }} 
    &=\left[1-\left(\textbf{\textsf{U}}_{\mathrm{f}}+\boldsymbol{\Delta}\right) \textbf{\textsf{S}}\left(\textbf{\textsf{U}}_{\mathrm{f}}+\boldsymbol{\Delta}\right)^\text{T}\right]\left(\textbf{\textsf{X}}_{\mathrm{f}}+\textbf{\textsf{X}}_{\mathrm{s}}\right) \\
    & = \textbf{\textsf{X}}_\text{f}+\textbf{\textsf{X}}_\text{s}- \textbf{\textsf{U}}_{\mathrm{f}}\textbf{\textsf{S}}\textbf{\textsf{U}}_{\mathrm{f}}^\text{T}\textbf{\textsf{X}}_\text{f}- \textbf{\textsf{U}}_{\mathrm{f}}\textbf{\textsf{S}}\textbf{\textsf{U}}_{\mathrm{f}}^\text{T}\textbf{\textsf{X}}_\text{s} - \Delta \textbf{\textsf{S}} \textbf{\textsf{U}}_{\mathrm{f}} \textbf{\textsf{X}}_{\mathrm{f}}-\textbf{\textsf{U}}_{\mathrm{f}} \textbf{\textsf{S}} \Delta^\text{T} \textbf{\textsf{X}}_{\mathrm{f}}\\ 
    & \quad-\boldsymbol{\Delta}\textbf{\textsf{S}}\boldsymbol{\Delta}^\text{T}\textbf{\textsf{X}}_\text{s}-\boldsymbol{\Delta}\textbf{\textsf{S}}\boldsymbol{\Delta}^\text{T}\textbf{\textsf{X}}_\text{f}-\boldsymbol{\Delta}\textbf{\textsf{S}}\textbf{\textsf{U}}_\text{f}^\text{T}\textbf{\textsf{X}}_\text{s}-\textbf{\textsf{U}}_\text{f}\textbf{\textsf{S}}\boldsymbol{\Delta}^\text{T}\textbf{\textsf{X}}_\text{s}\,.
\end{aligned}\label{eq:Xclean}
\end{equation}
In \autoref{fig:DecomposedTerms} we show power spectra for the subtracted decomposed terms in \autoref{eq:Xclean}, plotting their cross-correlation with the pure \hi\ signal to demonstrate where signal loss originates. Since these are the subtracted terms, the higher their cross-power with pure-\hi, the more they are contributing to signal loss. For reference, we also show the pure-\hi\ (i.e. the \hi\ auto-correlation) as the black-dotted line, and the fully cleaned result (all terms from \autoref{eq:Xclean} combined) as the grey dotted line.

As expected, a large bulk of signal loss is caused by the subtraction of the $\textbf{\textsf{U}}_\text{f}\textbf{\textsf{S}}\textbf{\textsf{U}}_\text{f}^\text{T}\textbf{\textsf{X}}_\text{s}$ term (blue dashed line). This \textit{direct} signal loss will clearly increase for higher $\Nfg$ i.e. more ones along the diagonal of $\textbf{\textsf{S}}$, and this is demonstrated by the growing amplitude of the blue dashed line mostly at small-$k$, going left to right in the panels. The projected out foregrounds $\textbf{\textsf{U}}_\text{f}\textbf{\textsf{S}}\textbf{\textsf{U}}_\text{f}^\text{T}\textbf{\textsf{X}}_\text{f}$ are entirely uncorrelated with the \hi\ signal as shown by the consistent with zero power spectrum (orange line). However, we still decrease the amplitude of $\textbf{\textsf{U}}_\text{f}\textbf{\textsf{S}}\textbf{\textsf{U}}_\text{f}^\text{T}\textbf{\textsf{X}}_\text{f}$ by three orders of magnitude (indicated in the legend) since this dominant term still has large purely statistical fluctuations in the \hi\ cross-correlation dependent only on the foreground realisation. The thinner lines represent terms including perturbative contributions from the \hi\ signal $\mathbf{\Delta}$. This is where the issue of signal loss begins to complicate. As $\Nfg$ increases, the contribution to signal loss from $\mathbf{\Delta}\textbf{\textsf{S}}\mathbf{\Delta}^\text{T}\textbf{\textsf{X}}_\text{s}$ (brown dashed line) becomes non-negligible. Complicating matters further is the removal of the anti-correlating contribution in $\mathbf{\Delta}\textbf{\textsf{S}}\textbf{\textsf{U}}^\text{T}_\text{f}\textbf{\textsf{X}}_\text{s}$. Lastly, there are also noticeable correlations in the perturbed foreground removed terms. The thin red line shows how the removed term $\mathbf{\Delta}\textbf{\textsf{S}}\textbf{\textsf{U}}^\text{T}_\text{f}\textbf{\textsf{X}}_\text{f}$ will also introduce a contribution to signal loss. This explains the previous result presented in the bottom-left panel of \autoref{fig:Xclean_comb} where, despite only projecting out modes from pure foreground data $\textbf{\textsf{X}}_\text{f}$, signal loss still arose in the cleaned power spectrum, albeit at a ${\sim}\,5\%$ level in the largest scales. This will be caused by the signal perturbations $\mathbf{\Delta}$ which introduce a small correlation with the \hi\ signal, shown by the red line in \autoref{fig:DecomposedTerms}. 

The complex mix of signal correlation and anti-correlation caused by the perturbations $\mathbf{\Delta}$ from non-foreground modes can clearly affect the overall signal loss. The impact of signal perturbations on the foreground modes becomes increasingly more important the more aggressive the foreground clean due to $\mathbf{\Delta}$ having more influence over less dominant foreground modes, as seen in \autoref{fig:PerturbedEigenvecs}. It is therefore crucial to model or emulate all these contributions in any signal reconstruction to avoid unbiased results as highlighted in previous literature \citep{Switzer:2015ria,Cheng:2018osq}. In \secref{sec:TF} we will explore how signal injection can be used to construct a foreground transfer function and validate with simulations how it is able to successfully emulate all the subtle contributions to signal loss. We will also explicitly highlight cases where a transfer function can be incorrectly assembled such that some of the contributions demonstrated by \autoref{fig:DecomposedTerms} are not accounted for, leading to incorrect estimations of signal loss.

\section{Validating the transfer function with simulations}\label{sec:TF}

As demonstrated in the previous section, signal loss from blind foreground cleaning is complicated by the subtraction of sub-sets of data that have spurious correlations and anti-correlations with the \hi\ signal. The spurious correlations arise because the estimated modes projected out in the blind foreground clean are perturbed by the presence of the \hi\ signal itself. Thus the signal loss is dependent on the specific realisation of the signal, foregrounds, and their combination. Modelling the signal loss, or measuring it in pure simulations, is therefore potentially problematic and could lead to biased results. In this section, we explore how we can utilise the observed contaminated data itself to emulate the complex spurious correlations in injected mock data and use the signal loss experienced in the mocks to construct a foreground transfer function. This data-driven approach has been extensively used in single-dish intensity mapping detections \citep{Masui:2012zc,Anderson:2017ert,Wolz:2021ofa,Cunnington:2022uzo}. Here we present the formalism for an unbiased application of the transfer function and for the first time validate its performance on simulations whilst also trying to expose any limitations.

Throughout this section, we treat the MD1GPC simulation as the \textit{observed} intensity mapping data, and use lognormal mocks as completely separate simulations in the construction of the transfer functions. This maintains a certain independence between the injected mocks and the simulated signal that we are trying to recover, as would be the case in real observations. There is also the option of generating more complex mocks to inject into the data, for example using field level forward modelling \citep{Obuljen:2022cjo} or a \hi-halo prescription as trialed in \citet{Wolz:2021ofa}, however we found lognormal mocks sufficient for our purposes.

In this section, in cases where we are investigating the cleaned or reconstructed power spectrum, unless stated otherwise, we will use the cross-correlation power spectrum between the foreground cleaned MD1GPC map and the \hi-only (foreground-free) MD1GPC map. This is so foreground residuals will be less of an issue and their additive bias does not confuse the investigation of signal loss and reconstruction accuracy. In cross-correlation with the \hi-only map, any difference relative to the original \hi-only auto-power spectrum will be caused solely by signal loss.

\subsection{Summarised recipe for the transfer function and its unbiased results}\label{sec:TFsummary}

We begin by providing a summary of how 
a transfer function can be used for correcting signal loss from foreground cleaning, and validate the performance of the process. We then go into more detail in the remainder of this section, clarifying exactly how the transfer function can be constructed and used for various scenarios. In short, the foreground transfer function is constructed by injecting mock data into the observed maps. Then, by running the same foreground removal routine, one will subject the mock data to a similar signal loss that is experienced by the true underlying \hi\ signal, thus giving an estimate of the true signal loss. 

Below is the step-by-step recipe for how to construct and apply an unbiased transfer function;

\begin{enumerate}[wide, labelindent=0pt,label=\textbf{(\roman*)}]
    \item Firstly, foreground clean the observed data $\textbf{\textsf{X}}_\text{obs}$ by projecting out $\Nfg$ PCA modes i.e. $\textbf{\textsf{X}}_\text{clean}\txteq{=}\textbf{\textsf{X}}_\text{obs}\txteq{-}\textbf{\textsf{U}}\textbf{\textsf{S}}\textbf{\textsf{U}}^\text{T}\textbf{\textsf{X}}_\text{obs}$, where $\textbf{\textsf{U}}$ is a matrix of eigenvectors from the diagonalisation of the $\nu,\nu'$ covariance matrix estimated empirically from the data, and $\textbf{\textsf{S}}$ is the selection matrix with ones along the first $\Nfg$ diagonal elements and zeros elsewhere. 
    \\
    \item Compute the power spectrum for the foreground-cleaned data, which will be negatively biased due to the signal loss from the foreground clean. The power spectrum is given here by $P_\text{clean}(k)\txteq{=}\mathcal{P}(\textbf{\textsf{X}}_\text{clean},\textbf{\textsf{X}}_\text{tr})$, where $\mathcal{P}(\textbf{\textsf{X}}_\text{clean},\textbf{\textsf{X}}_\text{tr})$ is an operator which measures the cross-power spectrum between data sets $\textbf{\textsf{X}}_\text{clean}$ and $\textbf{\textsf{X}}_\text{tr}$, then reduces the power into the spherically-averaged $k$-bins. Here, $\textbf{\textsf{X}}_\text{tr}$ is any overlapping tracer, which can be the intensity map itself for an auto-correlation survey, or a galaxy survey as a common example of cross-correlation.
    \\
    \item Generate mock \hi\ signal maps $\textbf{\textsf{X}}_\text{m}$ with the same dimensions as the observed data. In this work we use a fast lognormal transform process to generate mocks from the \hi\ power spectrum model given in \autoref{eq:Pkmodel}. We investigate the consequences of variation in the input mocks in \secref{sec:cosmo}.
    \\
    \item Emulate signal loss in the mock by injecting it into the real data and foreground cleaning the observed data and mock combination,
    \begin{equation}\label{eq:Mclean}
        \textbf{\textsf{X}}^\text{m}_\text{clean} \txteq{=}(\textbf{\textsf{X}}_\text{obs}\txteq{+}\textbf{\textsf{X}}_\text{m})\txteq{-} \textbf{\textsf{U}}_\text{obs+m}\textbf{\textsf{S}}\textbf{\textsf{U}}_\text{obs+m}^\text{T}(\textbf{\textsf{X}}_\text{obs}\txteq{+}\textbf{\textsf{X}}_\text{m})\txteq{-} [\textbf{\textsf{X}}_\text{clean}]\,.
    \end{equation}
    The term in the square brackets is subtracting the cleaned observed data without mocks to remove contributions in the map uncorrelated to the mock signal thus reducing the variance of the transfer function, as we will explicitly demonstrate.
    \\
    \item The foreground transfer function is then given by
    \begin{equation}\label{eq:TransferFunc}
	    \mathcal{T}(k) = \left\langle  \frac{\mathcal{P}(\textbf{\textsf{X}}^\text{m}_\text{clean}\, ,\, \textbf{\textsf{X}}_\text{m})}{\mathcal{P}(\textbf{\textsf{X}}_\text{m}\, ,\, \textbf{\textsf{X}}_\text{m})} \right\rangle_{N_\text{mock}} \,,
    \end{equation}
    where the angled brackets denote an averaging over iterations of a suitably large number of mocks ($N_\text{mock}$) until a converged transfer function is achieved. We use $N_\text{mock}\txteq{=}100$ by default unless otherwise mentioned.
    \\
    \item De-bias the cleaned power spectrum  using the transfer function to reconstruct the signal loss with $P_\text{rec}(k)\txteq{=}P_\text{clean}(k)[\mathcal{T}(k)]^{-1}$. Note the index of $-1$ should also be used in auto-correlation i.e. an auto-correlation of an intensity map should \textit{not} have signal loss corrected for twice, as we will demonstrate in \secref{sec:autovcross}.\\

    \item The covariance of the reconstructed power spectrum can also be extracted from the mocks used in the transfer function calculation. Whilst the mean of $P_\text{clean}\mathcal{T}^{-1}_i$ over all $N_\text{mock}$ iterations provides the reconstructed power spectrum, the covariance estimates the errors inclusive of signal loss uncertainty. However, as we will show, it is crucial not to subtract the square-bracket $\textbf{\textsf{X}}_\text{clean}$ term in \autoref{eq:Mclean} when estimating the covariance, as this will include foreground residuals, instrumental noise, etc. all of which should contribute to the error budget. We discuss this in detail in \secref{sec:ErrorEst}.
\end{enumerate}

\noindent The numerator in \autoref{eq:TransferFunc} is taking the cross-correlation between the cleaned mock $\textbf{\textsf{X}}^\text{m}_\text{clean}$ and original mock $\textbf{\textsf{X}}_\text{m}$ with no cleaning effects. This should therefore not be overly influenced by foreground residuals and differences between this cross-correlation and the auto-correlation in the denominator should only be caused by signal loss from the foreground clean, thus their ratio provides the level of the original signal remaining in the power spectrum of the cleaned mock $\textbf{\textsf{X}}^\text{m}_\text{clean}$. The crucial part for \autoref{eq:TransferFunc} is having a process for obtaining $\textbf{\textsf{X}}^\text{m}_\text{clean}$ such that the signal loss it experiences across all scales, is the same as the signal loss in the actual data $\textbf{\textsf{X}}_\text{s}$. To achieve this we inject mock signal $\textbf{\textsf{X}}_\text{m}$ into the observed data with foregrounds and true \hi\ signal ($\textbf{\textsf{X}}_\text{f+s+m}\txteq{\equiv}\textbf{\textsf{X}}_\text{f}\txteq{+}\textbf{\textsf{X}}_\text{s}\txteq{+}\textbf{\textsf{X}}_\text{m}$) then project out the same number of modes as in the original foreground clean of the observations i.e.;
\begin{equation}\label{eq:Mcleantext}
    \textbf{\textsf{X}}^\text{m}_\text{clean} =\textbf{\textsf{X}}_\text{f+s+m} - \textbf{\textsf{U}}_\text{f+s+m}\textbf{\textsf{S}}\textbf{\textsf{U}}_\text{f+s+m}^\text{T}\textbf{\textsf{X}}_\text{f+s+m} - [\textbf{\textsf{X}}_\text{clean}]\,. 
\end{equation}
This is equivalent to what we presented in the summarised recipe in \autoref{eq:Mclean}. 
As discussed, the term in the square bracket is subtracting the cleaned observed data (with no mock injection) to reduce the transfer function variance, which we discuss in more detail later. The presence of mock signal will cause perturbations to the eigenmodes, and will emulate the signal loss coming from both projecting out the modes with signal perturbations and the complex correlations between all the cross terms discussed in \secref{sec:OriginsSubtletiesSignalLoss} and \autoref{eq:Xclean}.

The presence of the true observed \hi\ signal $\textbf{\textsf{X}}_\text{s}$ in $\textbf{\textsf{X}}_\text{f+s+m}\txteq{\equiv}\textbf{\textsf{X}}_\text{f}\txteq{+}\textbf{\textsf{X}}_\text{s}\txteq{+}\textbf{\textsf{X}}_\text{m}$ in \autoref{eq:Mcleantext} creates unwanted complications to the transfer function construction which should ideally only be concerned with the mock signal $\textbf{\textsf{X}}_\text{m}$ and its relationship with $\textbf{\textsf{X}}_\text{f}$. The presence of $\textbf{\textsf{X}}_\text{s}$ will not matter from a \textit{direct} signal loss perspective, since this will not affect the cross-correlation with $\textbf{\textsf{X}}_\text{m}$ in \autoref{eq:TransferFunc}. However, $\textbf{\textsf{X}}_\text{s}$ will perturb the estimation of the eigenvectors. This is unwanted because we ideally only want the mock signal to perturb the eigenvectors and just produce $\textbf{\textsf{U}}_\text{f+m}$, but by injecting mocks into the true signal we will actually measure 
\begin{equation}\label{eq:EigenmodeWithMock+SignalPertub}
    \textbf{\textsf{U}}_\text{f+s+m} = \textbf{\textsf{U}}_\text{f} + \boldsymbol{\Delta}_\text{s} + \boldsymbol{\Delta}_\text{m}\,,
\end{equation}
where we have introduced the subscripts m and s to the perturbations $\mathbf{\Delta}$ to distinguish perturbations from mock signal and true \hi\ signal respectively. The two sources of perturbation is not seen in the foreground clean of just the observations ($\textbf{\textsf{X}}_\text{f}\txteq{+}\textbf{\textsf{X}}_\text{s}$). In other words the eigenvectors are now being perturbed twice. As we will show from our results shortly, this appears to have little impact and we still obtain an unbiased transfer function. We tested the transfer function in an idealised case where mock signal was injected into just pure foreground ($\textbf{\textsf{X}}_\text{f}\txteq{+}\textbf{\textsf{X}}_\text{m}$) and found little difference in performance compared to the realistic case where true signal is present ($\textbf{\textsf{X}}_\text{f}\txteq{+}\textbf{\textsf{X}}_\text{s}\txteq{+}\textbf{\textsf{X}}_\text{m}$).

\begin{figure}
    \centering
    \includegraphics[width=1\linewidth]{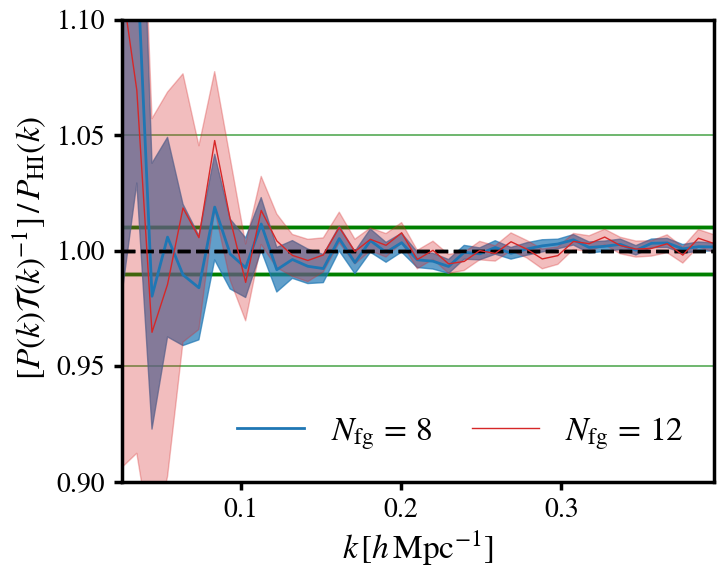}
    \caption{Accuracy of reconstructed foreground cleaned power spectra relative to the foreground-free \hi-only data ($P_\hi$) from simulated intensity maps. The foreground cleaned power spectrum has been reconstructed using the transfer function to correct for the signal loss from foreground cleaning. The transfer functions are calculated using \autoref{eq:TransferFunc} and \ref{eq:Mcleantext} and averaging over 100 lognormal mocks. The shaded bands show the rms over these 100 mocks. Results for a mild ($\Nfg\txteq{=}8$, blue lines) and more aggressive ($\Nfg\txteq{=}12$, red lines) foreground cleans are shown. \textcolor{black}{Dark-thick (light-thin) green horizontal lines indicate sub 1\% (5\%) accuracy regions of the reconstructed power spectrum.}}
    \label{fig:TFdifferentcases}
\end{figure}

The accuracy of the reconstructed power spectra is demonstrated in \autoref{fig:TFdifferentcases}. The simulated observations are cleaned by removing either $\Nfg\txteq{=}8$ or $12$ PCA modes, then the transfer function is used to correct for the signal loss, with the reconstructed result being divided by the original foreground-free simulation $P_\hi(k)$. Thus, a perfect reconstruction would give unity across all scales. We see excellent performance with sub-percent accuracy achieved across most scales above $k\txteq{>}0.1\hMpc$ for the $\Nfg\txteq{=}8$ case. Performance is still good for the $\Nfg\txteq{=}12$ case, albeit with a noticeable drop in accuracy relative to $\Nfg\txteq{=}8$ mostly at large scales (small-$k$). This will be caused by the increased effect from spurious correlations between foregrounds and mock signal, which, as we demonstrated in \autoref{fig:DecomposedTerms}, increases for more aggressive (higher $\Nfg$) foreground cleans. This will not necessarily bias the results since the variance in the transfer function also increases for higher $\Nfg$, as shown by the shaded regions, thus can be reflected in the error estimations (discussed in a later section). 

In general on large ($k\txteq{<}0.1\hMpc$) scales, we see a less reliable result in terms of pure accuracy, but this is also accounted for by the transfer function variance, which can reach ${\gtrsim}\,5\%$ on these scales. The performance at large scales is however dependent on the size of the intensity mapping survey. The depth of the $1\,(\text{Gpc}/h)^3$ MD1GPC simulation at $z\txteq{\sim}0.39$ is reasonably consistent with a MeerKAT L-band survey, assuming it uses the complete band range ($0.2\txteq{<}z\txteq{<}0.58$). However, future surveys in UHF band, and then eventually the SKAO, will cover much wider frequency ranges. This will mean reconstructed modes at $k\txteq{<}0.1\hMpc$ become more reliable due to a suppression of sample variance and less signal loss which will now be contained to even larger scales. We will demonstrate this point later in \secref{sec:cosmo} with some additional simulations which cover a larger volume. 

The presence of the true observed \hi\ signal in the transfer function calculation will increase its variance because there will be residual true signal after the foreground clean. This will be uncorrelated from the mocks and act like noise and increase the variance across all of the mocks being averaged over. This is why we subtract the cleaned data (the $\textbf{\textsf{X}}_\text{clean}$ term in the square brackets of \autoref{eq:Mcleantext}), since this is only contributing variance to the result. We will revisit this point in the next section where we will demonstrate that the increased variance coming from $\textbf{\textsf{X}}_\text{clean}$ can be utilised for error estimation. With the $\textbf{\textsf{X}}_\text{clean}$ subtraction, this version of the transfer function is not only achieving a good accuracy but also a good uncertainty on most scales, shown by the shaded region.

The validation of the transfer function demonstrated by \autoref{fig:TFdifferentcases} is an important result. This is a method for reconstructing signal loss which is applicable on real data and delivers unbiased results across all scales and within sub-percent precision across smaller scales where the particular survey volume allows those modes to be well sampled ($k\txteq{>}0.1\hMpc$ for the case of the MD1GPC simulations). The compromise of having to inject mock signal into a combination of both foreground and true \hi\ signal is an unavoidable complication, however, there is no evidence that this causes any bias in the reconstructed power spectrum. Furthermore, by subtracting the cleaned data ($\textbf{\textsf{X}}_\text{clean}$ term in the square brackets of \autoref{eq:Mclean}), we found the increase in variance relative to an ideal case where no true signal ($\textbf{\textsf{X}}_\text{s}$) is present in the transfer function calculation was only ${\sim}\,20\%$. It is crucial that the form of the transfer function as defined by \autoref{eq:TransferFunc} and \autoref{eq:Mclean} be followed and in \appref{sec:BiasedTF} we explicitly highlight the consequences of deviating from this prescription, demonstrating the significant biases caused when different definitions of the transfer function are used. 

\begin{figure}
    \centering
    \includegraphics[width=1\linewidth]{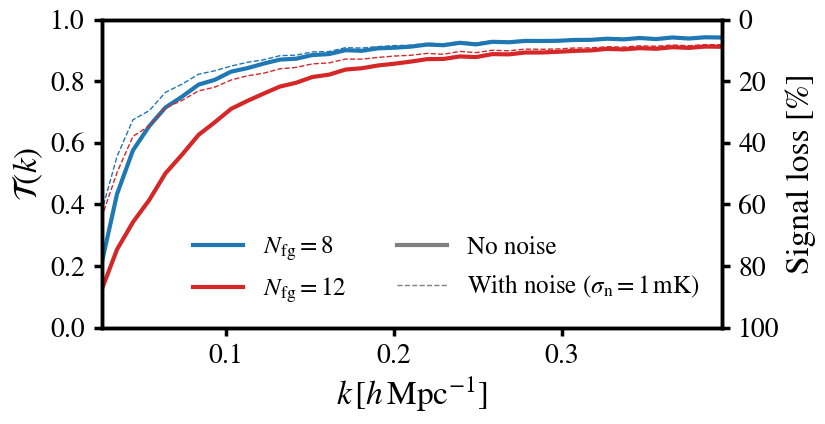}
    \caption{Foreground transfer functions $\mathcal{T}(k)$ constructed using \autoref{eq:TransferFunc} and \autoref{eq:Mclean} for the MD1GPC simulations for different numbers of PCA modes removed (given by $\Nfg$). Solid lines indicate noise-free simulations, thin dashed-lines are for cases where white noise with rms $\sigma_\text{n}\txteq{=}1\,\text{mK}$, which dominates over the \hi, is added to the simulated observations.}
    \label{fig:TFdemo1D}
    \includegraphics[width=0.9\linewidth]{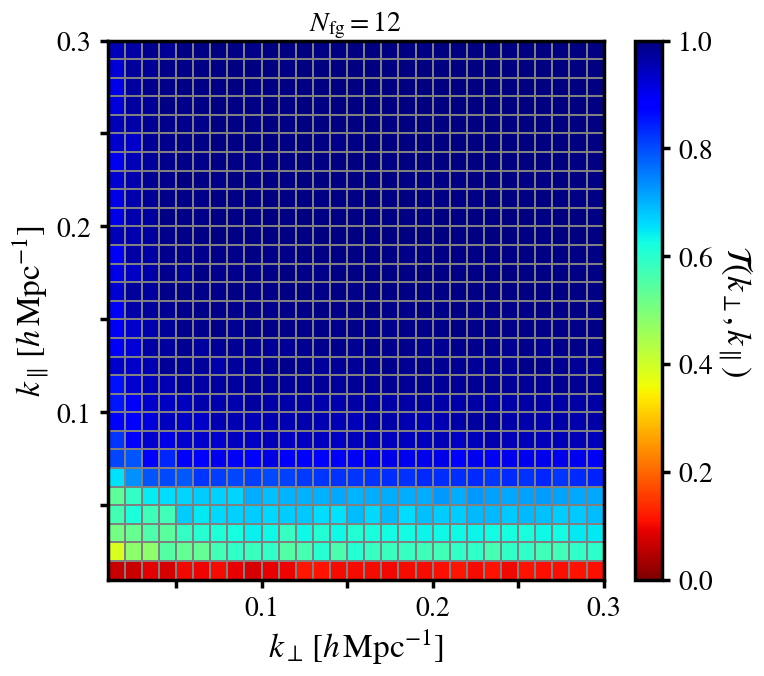}
    \caption{Similar to \autoref{fig:TFdemo1D}, this shows the foreground transfer function but now in cylindrical $k_\perp,k_\parallel$ space where $\Nfg\txteq{=}12$ PCA modes have been removed. This is for the noise-free case.}
    \label{fig:TFdemo2D}
\end{figure}

In \autoref{fig:TFdemo1D} and \autoref{fig:TFdemo2D} we demonstrate the \textit{shape} of the transfer function in $k$-space and in doing so, analyse where signal loss is most severe. \autoref{fig:TFdemo1D} shows transfer functions for different PCA modes removed (given by $\Nfg$). This confirms that signal loss increases with $\Nfg$ and is higher at smaller-$k$, both as expected. We also show the impact from adding the dominant instrumental noise. Perhaps counter-intuitively, this causes less signal loss. This is because the noise is the dominant source of perturbations to the pure foreground modes (as shown by \autoref{fig:PerturbedEigenvecs}), hence these noise-dominant modes will have less contribution from the \hi\ signal and removing them causes less signal loss. However, this will result in a poorer overall foreground clean. Thus foreground residuals and the high-level noise already present will cause problems for additive biases in auto-correlation and would also lead to higher errors in cross-correlation. So the presence of high noise is not beneficial as \autoref{fig:TFdemo1D} alone may appear to suggest. We explore the high noise scenario further in \secref{sec:MK} where we use simulations which emulate an early pathfinder-like intensity mapping survey.

Similarly to \autoref{fig:TFdemo1D}, an example transfer function is also shown in \autoref{fig:TFdemo2D} but now decomposed into cylindrical contributions in $k_\perp,k_\parallel$. As already well established, signal loss is overwhelmingly a function of small-$k_\parallel$. However, there is also some slight $k_\perp$ dependence with signal loss being slightly higher at small-$k_\perp$ caused by the large angular structures in the foregrounds. It is important to highlight that the nature of signal loss will vary depending on not just the foreground's strength and spectral smoothness, but also on the depth of the survey in frequency. This means that the signal loss presented in \autoref{fig:TFdemo1D} and \autoref{fig:TFdemo2D} is specific to the MD1GPC simulation. However, the conclusions we have drawn from this are still mostly generic. For example, signal loss is still contained in small-$k_\parallel$ modes in more systematic dominated intensity maps as shown in recent MeerKAT analysis \citep{Cunnington:2022uzo} and as we will show later in the simulations designed to emulate a small MeerKAT pilot intensity mapping survey. Whilst signal loss appears widespread throughout all modes in the spherically averaged \autoref{fig:TFdemo1D}, where ${>}\,5\%$ signal loss is evident even on small scales, it is clear from \autoref{fig:TFdemo2D} that signal loss does tend to zero (where $\mathcal{T}(k_\perp,k_\parallel)\txteq{\sim}1$) above some $k_\parallel$ cut. This raises an intriguing possibility of adopting a hybrid foreground cleaning/avoidance strategy where a blind foreground clean is run on the full data, but then an avoidance strategy is used where only modes above some $k_\parallel$ are kept for further analysis. At the very least this would limit the dependence on the transfer function but would unlikely be reliable enough to completely avoid any use of signal reconstruction. Furthermore, the methodology of the transfer function would still be a required tool for robustly assessing where an optimum cut in $k_\parallel$ should be made. Scale cuts will also limit the scope and constraints possible with the experiment. We defer any investigations of $k_\parallel$ cuts to future work and continue to test the transfer function on small-$k_\parallel$, especially since these are the scales that will test the performance of the transfer function most stringently.

\subsection{1D vs 2D bandpowers in the transfer function}\label{sec:2DTF}

Our results so far have reduced all power directly into 1D spherically averaged $k$-bins and the results in \autoref{fig:TFdifferentcases} suggest this can be sufficient. Providing the same $k$-bins are used in the 1D spherical averaging for the transfer function construction and cleaned power spectrum, then the transfer function should encapsulate the same anisotropic signal loss in each $k$-bin as inflicted on the cleaned data. Furthermore, going straight to 1D $k$-bins avoids extra compression steps which could potentially lead to results being lossier. However, it has been proposed in the literature that because the signal loss is anisotropic (demonstrated by \autoref{fig:TFdemo2D}) that the transfer function should be estimated and applied in 2D cylindrical $k_\perp, k_\parallel$ space, with these bandpowers then re-binned to provide the final spherically averaged 1D power spectrum \citep{Masui:2012zc,Switzer:2015ria}.

\begin{figure}
    \centering
    \includegraphics[width=1\linewidth]{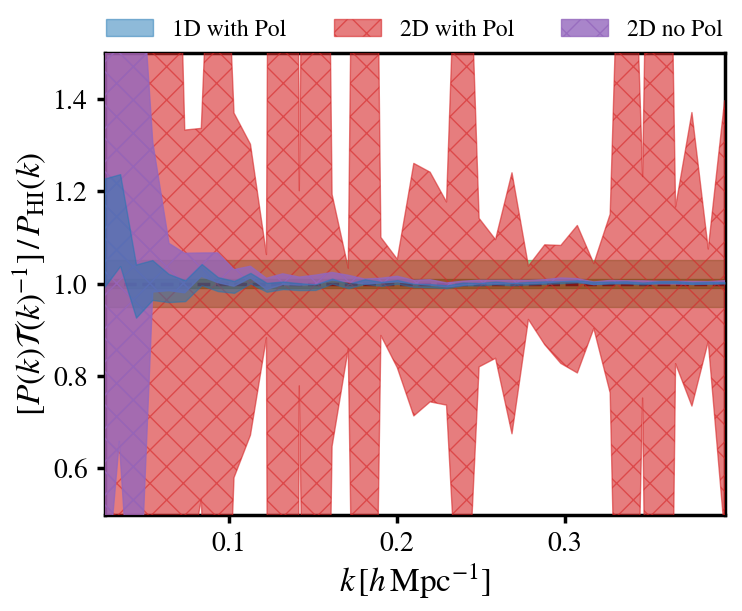}
    \caption{Demonstration of the large increase in the variance of the reconstructed power spectrum, when the transfer function is constructed and applied in 2D $k_\perp,k_\parallel$ space. Blue results show the standard 1D cases (same as \autoref{fig:TFdifferentcases}) with $\Nfg\txteq{=}8$. Red-hatched results show the 2D construction following steps in \secref{sec:2DTF}. Purple-hatched results also show results with a 2D construction, but with simplified foregrounds excluding the polarisation leakage.}
    \label{fig:AnisotropicTF}
\end{figure}

We tested the 2D-cylindrical transfer function approach and found evidence of higher variance in the standard case where complicated polarised foregrounds are present in the observations. We demonstrate this in \autoref{fig:AnisotropicTF}. Here, the 1D reconstruction (blue shading) shows our default setup used everywhere else in the paper, averaging straight into 1D spherical $k$-bins. The 2D reconstruction refers to a case where we average all power into $100\, k_\perp \times 100\,k_\parallel$ cylindrical linear-spaced bins, with $0\txteq{<}k_\perp,k_\parallel\txteq{<}0.4\,h\,\text{Mpc}^{-1}$, in the transfer function construction. The measured power for the cleaned observations is also reduced into the same 2D bins and a reconstructed power spectrum for a single $i$th mock iteration is given by $P_{\text{rec},i}(k_\perp,k_\parallel)\txteq{=}P_\text{clean}(k_\perp,k_\parallel)/\mathcal{T}_i(k_\perp,k_\parallel)$. The 2D powers then undergo a weighted average into 1D $k$-bins to give the rebinned 1D power spectrum, defined by 
\begin{equation}\label{eq:2Dto1D}
    P_{\text{rec},i}(k) = \frac{\sum_{\alpha}N_\alpha P_{\text{rec},i}(k_{\perp,\alpha},k_{\parallel,\alpha})}{\sum_{\alpha}N_\alpha}\,,
\end{equation}
where all unique 2D $k_{\perp},k_{\parallel}$ bandpowers are indexed by $\alpha$ and the summation is over all 2D powers contained in the 1D bin $k\txteq{\equiv}\sqrt{k_\perp^2\txteq{+}k_\parallel^2}\txteq{\in}(k\txteq{-}\Delta k/2, k\txteq{+}\Delta k/2)$. $N_\alpha$ is the number of 3D Fourier modes contained in the particular 2D $k_{\perp,\alpha},k_{\parallel,\alpha}$ bandpower. Similar to the direct 1D reconstruction, the mean over all $i$th mocks in $P_{\text{rec},i}(k)$ gives the final estimated reconstructed power spectrum, and the variance provides an estimate of the expected errors. The results for the 2D rebinned power are shown by the red-hatched shading in \autoref{fig:AnisotropicTF} where the large increase in variance is clear. We found by switching off the complexity caused to the foregrounds by the simulated polarisation leakage made the 2D transfer function more reliable (purple hatched results), although still a higher variance is returned in these results, particularly at small-$k$, relative to the 1D case.

It is likely that outliers in isolated iterations are causing the large variance in \autoref{fig:AnisotropicTF}. These outliers would then be suppressed in the simpler case where there is no polarisation leakage, thus less complex residuals or signal loss to cause extreme spurious correlations in the mocks.  \textcolor{black}{In an attempt to suppress the variance, we increased the number of mocks used in the 2D polarisation leakage construction to 500, but this yielded no improvement.} An extension that may be necessary to avoid the blow-up in variance, is a more detailed weighting to the rebinning procedure we perform in \autoref{eq:2Dto1D}. In \citet{Switzer:2015ria} they discuss applying an inverse covariance weighting to maximize the 1D signal-to-noise ratio. This could down-weight some of the outliers in our iterations and suppress the large variance in the 2D polarised results. We defer this extension to future work, ideally with even more realistic sims where a conclusive study can be performed into whether the 2D transfer function construction can be more optimal. Since our results suggest that direct 1D construction is better performing, at least in the case of the spherically averaged power spectrum, we \textcolor{black}{use} this approach for the rest of the paper, unless presenting a 2D cylindrical transfer functions which we now only do for demonstration purposes. 

\subsection{Error estimation for power spectra reconstructed with a foreground transfer function}\label{sec:ErrorEst}

Evaluating how to correctly estimate the contributions to the error from foreground contamination and signal loss uncertainty will be crucial for future precision cosmology with \hi\ intensity mapping. This is the focus of this section. An approach taken in some previous intensity mapping detections \citep{Anderson:2017ert} has been to \textcolor{black}{use the variance over the mock simulations used in the transfer function construction} for the error estimate on cross-correlation measurements. \textcolor{black}{It is possible to capture this uncertainty from the variance in the transfer function i.e. 
the errors can be estimated using}
\begin{equation}\label{eq:varTFerr}
	\hat{\sigma}_{P_\text{c}} = \sigma\{P_\text{clean}\mathcal{T}^{-1}_i \}\, ,
\end{equation}
where $\mathcal{T}_i$ is the transfer function from the $i$th mock in the construction, and $\sigma\{\}$ is taking the rms over all $i$ iterations. The rms over all transfer function iterations which include the injected true observations should provide an error estimate which incorporates thermal noise, foreground residuals, residual RFI, sample variance and signal loss from foreground cleaning. \textcolor{black}{However, crucially this approach relies on modifying the transfer function definition so that the $\textbf{\textsf{X}}_\text{clean}$ term in \autoref{eq:Mclean} is not subtracted.}

\begin{figure}
    \centering
    \includegraphics[width=1\linewidth]{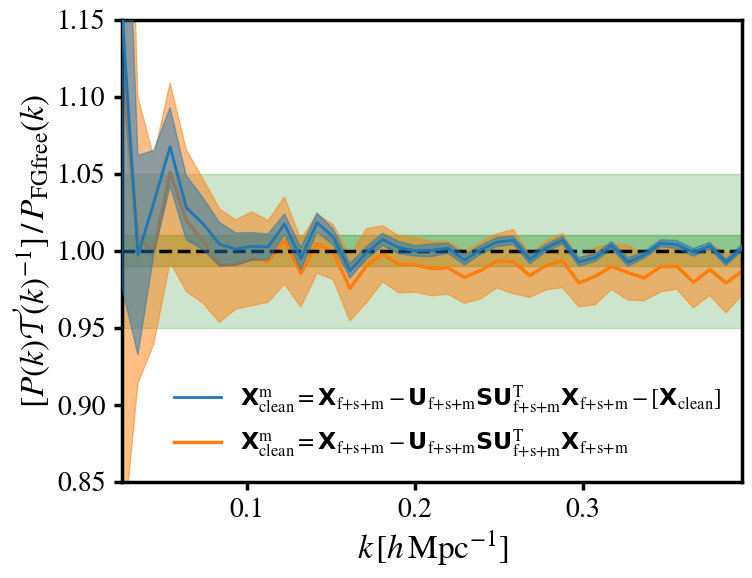}
    \caption{Impact from subtracting the cleaned observational data $\textbf{\textsf{X}}_\text{clean}$ in \autoref{eq:Mclean}. The shaded regions represent the rms over the 100 mocks used to construct the transfer function. The rms is significantly reduced when the cleaned data is subtracted. This test was done using a foreground clean with $\Nfg\txteq{=}8$ on simulations with high instrumental noise. Here the reconstructed result is divided by the foreground-free power spectrum $P_\text{FGfree}$ which contains identical instrumental noise.}
    \label{fig:TFwithnovarsubtractiondemo}
\end{figure}

We begin by demonstrating the impact subtracting the cleaned observed data $\textbf{\textsf{X}}_\text{clean}$ has on the transfer function. Since we are investigating error estimation, for this section it is helpful to use data with noticeable error-bar size, so we therefore add the dominant $\sigma_\text{n}\txteq{=}1\,\text{mK}$ noise to the MD1GPC simulations. \autoref{fig:TFwithnovarsubtractiondemo} shows the performance of the transfer function for the high-noise simulations, calculated using \autoref{eq:TransferFunc} and \autoref{eq:Mclean}, both with and without the $\textbf{\textsf{X}}_\text{clean}$ subtraction. It is encouraging to see that the addition of noise is not majorly affecting the performance of the transfer function. For the case where $\textbf{\textsf{X}}_\text{clean}$ has been subtracted (blue results), the accuracy is only mildly affected relative to the noise-free results in \autoref{fig:TFdifferentcases}. For the noise-inclusive results of \autoref{fig:TFwithnovarsubtractiondemo}, we divide by $P_\text{FGfree}$ in the $y$-axis which contains the same noise as the reconstructed power. This is to divide out the fluctuations caused by the presence of the dominant noise, allowing analysis into the performance of the reconstruction alone. For the case where $\textbf{\textsf{X}}_\text{clean}$ has not been subtracted (orange results), there is a slight drop in accuracy at small scales. This small bias is absent when we use a \hi\ signal without RSD thus it appears to be caused by the addition of uniform noise in the presence of an anisotropic \hi\ signal. We also found this small bias is decreased when we use the 2D transfer function construction outlined in \secref{sec:2DTF}, although this relied on there being no polarisation leakage which otherwise causes the variance of the result to blow up, as we showed. We discuss the performance of the transfer function in the presence of anisotropic phenomena later, but given this is a small bias and is absent in the subtracted $\textbf{\textsf{X}}_\text{clean}$ case, it is not overly important for this discussion on error covariance estimation. 

\autoref{fig:TFwithnovarsubtractiondemo} suggests that subtracting $\textbf{\textsf{X}}_\text{clean}$ is the favorable strategy if one is purely pursuing the most accurate transfer function possible. However, the main point of this plot is the difference in variance between the two cases. If one is using the variance on the transfer function as a basis for the error estimation, the reduced variance caused by subtracting $\textbf{\textsf{X}}_\text{clean}$ has implications and can lead to under-estimated errors on the reconstructed power spectrum, as we will now demonstrate.

To quantitatively evaluate error estimation performance, it is helpful to analyse how errors for a power spectrum measurement are analytically derived in a foreground-free case. The variance on a cross-correlation power spectrum $P_\text{c}$ between two tracers, 1 and 2, can be estimated as \citep{Feldman:1993ky}
\begin{equation}\label{eq:ErrorEst}
    \sigma_\text{theory}^2=\frac{1}{2 N_\text{m}} \left[P_\text{c}^2(k)+\left(P_1+\frac{V \sigma_1^2}{\langle f_1\rangle^2}\right)\left(P_2+\frac{V \sigma_2^2}{\langle f_2\rangle^2}\right)\right]\, ,
\end{equation}
where $N_\text{m}$ is the number of modes spherically averaged in each $k$-bin, $V$ is the volume of a single voxel on the Fourier grid, uncorrelated noise in the field is represented by the variance $\sigma^2$, which for an ideal intensity map would be the variance of the instrumental noise, and $\langle f\rangle$ is the background mean for the field e.g. the mean brightness temperature for intensity mapping. For galaxy surveys, the noise component given by the second terms in the curved brackets will reduce to shot-noise i.e. $V\sigma^2/\langle f\rangle^2\txteq{=}1/\bar{n}_\text{g}$, where $\bar{n}_\text{g}$ is the galaxy number density. We refer the reader to \citet{Blake:2019ddd} where a detailed derivation of the above is provided with applications to intensity mapping and its cross-correlations with galaxy surveys. 

Extending \autoref{eq:ErrorEst} to incorporate contributions from foreground contamination is challenging. Foreground residuals could presumably be estimated and would provide an additive variance, or be assumed sub-dominant enough not to warrant inclusion. However, the uncertainty from the transfer function, which for high signal loss is non-negligible at large scales, requires careful inclusion. The uncertainty in the transfer function can be estimated from the variance over the mocks used to construct it as we have shown, however, analytically adding this into the error budget of \autoref{eq:ErrorEst} is non-trivial since this will not necessarily be a contribution entirely independent of the noise and cosmic variance already being factored for in \autoref{eq:ErrorEst}. This is why some previous analyses have used the transfer function variance as a basis for overall error estimation. To evaluate whether this is a robust method, we can use the analytical errors as a benchmark. Errors estimated based on the variance in the transfer function should approximately agree on small scales with the analytical ones where foreground contamination and signal loss are minimal, but the large noise still dominates. 

We validated that the analytical errors (\autoref{eq:ErrorEst}) are a good estimate for the foreground-free case. Using the MD1GPC simulations with the large $\sigma_\text{n}\txteq{=}1\,\text{mK}$ Gaussian white-noise but with no foregrounds, we measure the cross-correlation power spectrum with a noise-free equivalent, then estimate the errors using \autoref{eq:ErrorEst} and evaluate the $\chi^2_\text{dof}$ given by 
\begin{equation}
	\chi^2_\text{dof} = \sum_k\frac{P_\text{data}(k)-P_\text{mod}(k)}{\sigma_P(k)} \Big/ (N_\text{k}-1)\, ,
\end{equation}
where $N_\text{k}$ is the number of $k$-bins. $P_\text{mod}$ is the model defined in \appref{sec:HIPkMod} with parameters matched to the \textsc{Multi-Dark} inputs used in the MD1GPC simulation. The analytical errors, return a $\chi^2_\text{dof}\txteq{\sim}1$ as expected, evidence that the errors are a reasonable size, given the reliable model. 

\begin{figure}
    \centering    \includegraphics[width=1\linewidth]{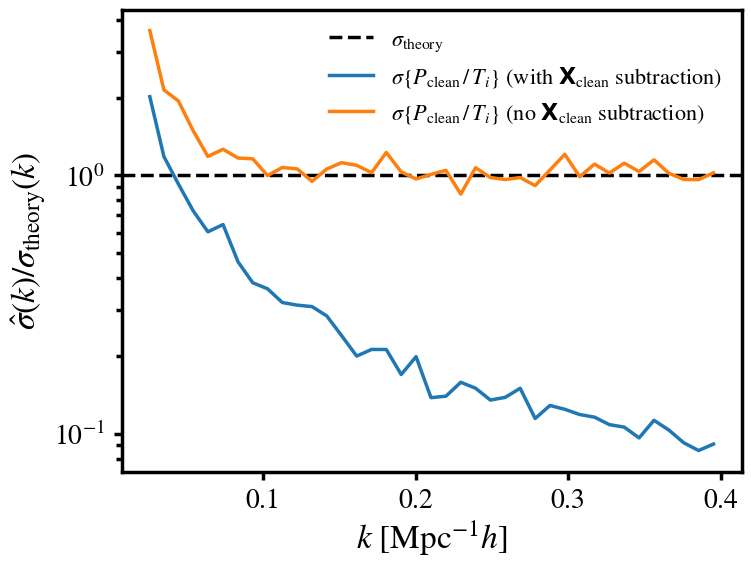}
    \caption{Comparison between methods of error estimation for simulations with signal loss reconstructed by a transfer function. The black dashed-line shows an analytical error estimate given by \autoref{eq:ErrorEst} which we have shown to be reliable for foreground-free data. The coloured lines show error estimation based on the variance of the transfer function, under two difference scenarios, with $\textbf{\textsf{X}}_\text{clean}$ subtraction (see \autoref{eq:Mclean}) and without. This is with the MD1GPC simulation where $\Nfg\txteq{=}8$ PCA modes removed.}
    \label{fig:ErrorEst}
    \includegraphics[width=1\linewidth]{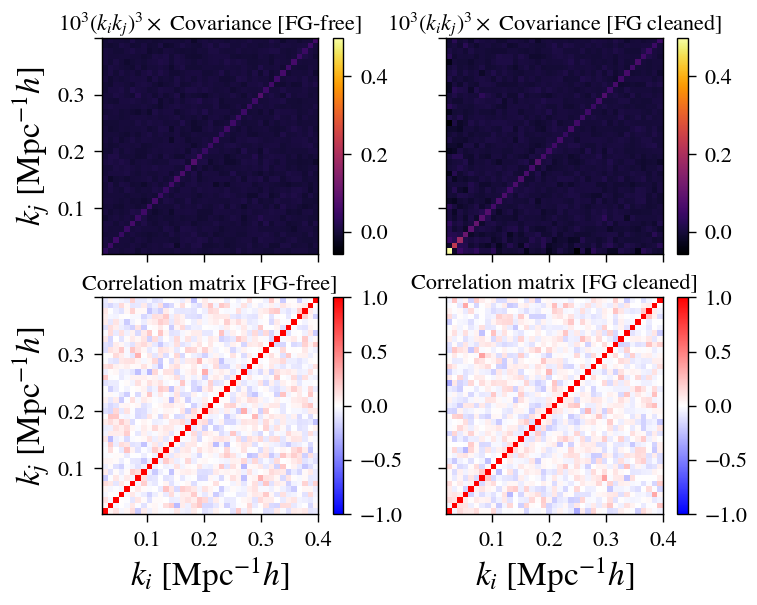}
    \caption{Covariance and correlation matrices for foreground-free (left column) and foreground cleaned with transfer function reconstructed signal loss (right column), without the $\textbf{\textsf{X}}_\text{clean}$ subtraction in the transfer function. Produced using the MD1GPC simulations with $\Nfg\txteq{=}8$ PCA modes removed, as in \autoref{fig:ErrorEst}. The covariance matrices have been multiplied by the product of the two-$k$-bins cubed times $10^3$ for demonstration purposes so high-$k$ covariance can be seen.}
    \label{fig:Covariance}
\end{figure}

Using the analytical errors $\sigma_\text{theory}$ as a validated benchmark, \autoref{fig:ErrorEst} shows how the error estimation based on the variance in the transfer function compares for the same simulations but with cleaned foregrounds. The blue line shows the case using a transfer function defined by \autoref{eq:TransferFunc} and \autoref{eq:Mclean}, where the cleaned observed data ($\textbf{\textsf{X}}_\text{clean}$) has been subtracted to reduce the variance. This is underestimating the errors relative to the analytical ones, likely caused by subtracting the $\textbf{\textsf{X}}_\text{clean}$ term which will contain noise and foreground residuals, which contribute to the error budget. \autoref{fig:ErrorEst} therefore shows that subtracting the cleaned data $\textbf{\textsf{X}}_\text{clean}$ in the transfer function construction results in the transfer function variance no longer being a reliable means for estimating the errors. However, it is clear from the orange line, which is equivalent to the blue but without the $\textbf{\textsf{X}}_\text{clean}$ subtraction, that the variance from this version of the transfer function leads to an excellent agreement with the analytical errors at high-$k$ as required. Furthermore, it also shows an increase in error at small-$k$ as one would expect where signal loss is highest, thus it is incorporating the increased uncertainty from signal reconstruction, not accounted for in the analytical errors.

Using the variance in the transfer function mocks for analysing uncertainties also has the advantage of being able to examine off-diagonal covariance of the data, something not trivially possible with the analytical approach. In \autoref{fig:Covariance} we show the $k_i k_j$ covariance matrix $C_{ij}$ (top row) as well as the normalised correlation matrix defined by $R_{ij}\txteq{=}C_{ij}/\sqrt{C_{ii}C_{jj}}$. The left column shows the foreground-free scenario where we inject mocks into the \hi\ + high-noise simulations to get an estimate of the covariance without a foreground cleaning step. The right column is equivalent but with cleaned foregrounds and a transfer function constructed without the subtraction of $\textbf{\textsf{X}}_\text{clean}$. The covariance over all iterations in the reconstructed power spectra (\autoref{eq:varTFerr}, but now including off-diagonal elements $i\txteq{\neq}j$) estimates the covariance of the observed data. As expected, and consistent with the orange line of \autoref{fig:ErrorEst}, the cleaned foregrounds are increasing covariance on large scales, but encouragingly they do not appear to increase off-diagonal correlations between $k$-bins. 

We conclude from this investigation that the variance in the transfer function is a reliable tool for error estimation in the final reconstructed power spectrum, provided the cleaned data $\textbf{\textsf{X}}_\text{clean}$ term is not subtracted in its construction. This becomes similar in approach to galaxy surveys which use vast suites of mocks as their primary method for estimating the covariance in their data \cite[e.g.][]{Zhao:2020bib}. If opting to use the transfer function variance for error estimation, it becomes important to ensure that all aspects of the error budget are emulated in the mocks used in the transfer function construction. An example of this would be in a galaxy survey cross-correlation where the galaxy shot noise would need to be captured by using galaxy mocks with the correct number densities and survey coverage. With the observational data injected into the mocks, we are also including variance from signal loss, foreground residuals, residual RFI, instrumental noise, etc. all of which are currently not well enough understood to reliably emulate in mock intensity maps\footnote{Jackknife resampling will also be a useful tool when unknown systematics are present.}.

\subsection{Auto- and cross-correlation applications}\label{sec:autovcross}

So far in this section, we have considered the cross-power spectrum between a foreground cleaned \hi\ intensity map and the original foreground-free, \hi-only map. This is so that any foreground residuals or noise in the cleaned maps do not complicate the analysis of \hi\ signal loss in the power spectra. Since foreground residuals and noise will not correlate with the \hi-only maps, the additive biases they cause are avoided in cross-correlation thus the only departure from the true-\hi\ power should be just from signal loss. However, \hi\ intensity mapping surveys will also aim to conduct analysis in auto-correlation and we need to consider how signal loss behaves in this scenario.

\begin{figure}
    \centering
    \includegraphics[width=0.9\linewidth]{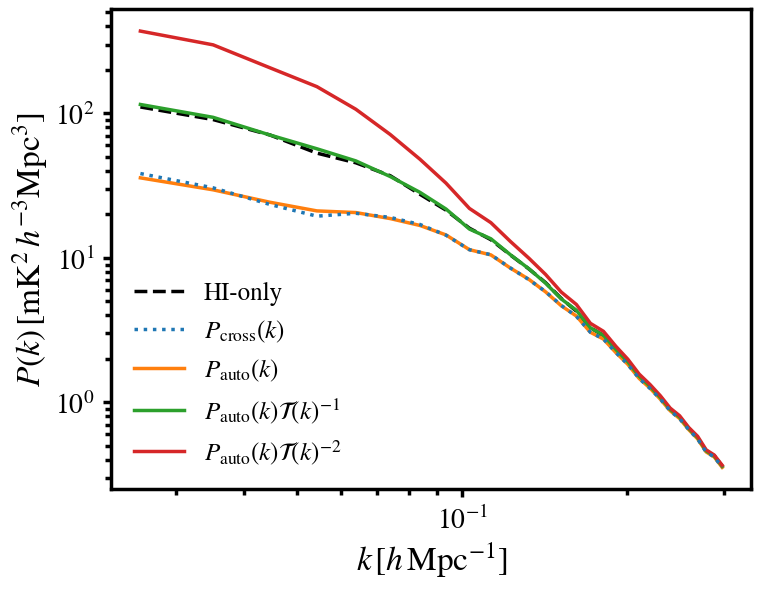}
    \caption{Demonstrating the correct application of the transfer function in an auto-correlation analysis. Black-dashed line shows the original foreground free power spectrum. The blue-dotted line shows the cross-correlation with the foreground-free. The orange line shows the auto-correlation. For this aggressive ($\Nfg\txteq{=}12$) foreground clean, foreground residuals should be minimal and the similar amplitude between $P_\text{auto}$ and $P_\text{cross}$ suggests the signal loss is similar in both. The green line shows the correct application of the transfer function and the red line shows the over-correction where $\mathcal{T}^{-2}$ is used.}
    \label{fig:AutoVsCross}
\end{figure}

It has been previously suggested that there would be twice as much signal lost in the auto-correlation power spectrum because its effects are present twice in the map product $P_\hi(\boldsymbol{k})\txteq{\propto}|\tilde{\textbf{\textsf{X}}}(\boldsymbol{k})|^2$. This would mean a $\mathcal{T}(k)^{-2}$ correction factor is needed to reconstruct the power spectrum. However, we found from our simulations that this is not the case, and the same degree of signal loss is also present in an auto-correlation as is in cross-correlation. In other words, the same correction of $\mathcal{T}(k)^{-1}$ is also needed in the auto-correlation as well as in cross-correlation. \autoref{fig:AutoVsCross} demonstrates this finding showing how the cross-correlation (blue-dotted line) and auto-correlation (orange-solid) appear to have approximately equivalent levels of signal loss. For this test, we return to the noise-free simulations and use an aggressive $\Nfg\txteq{=}12$ PCA clean which will suppress foreground residuals significantly, making it reasonable to ignore their influence on the results. The green line shows what appears to be the correct application of the $\mathcal{T}(k)^{-1}$ transfer function whereas the red line shows the consequential over-correction from applying the $\mathcal{T}(k)^{-2}$ to the auto-correlation.

\begin{figure}
    \centering
    \includegraphics[width=1\linewidth]{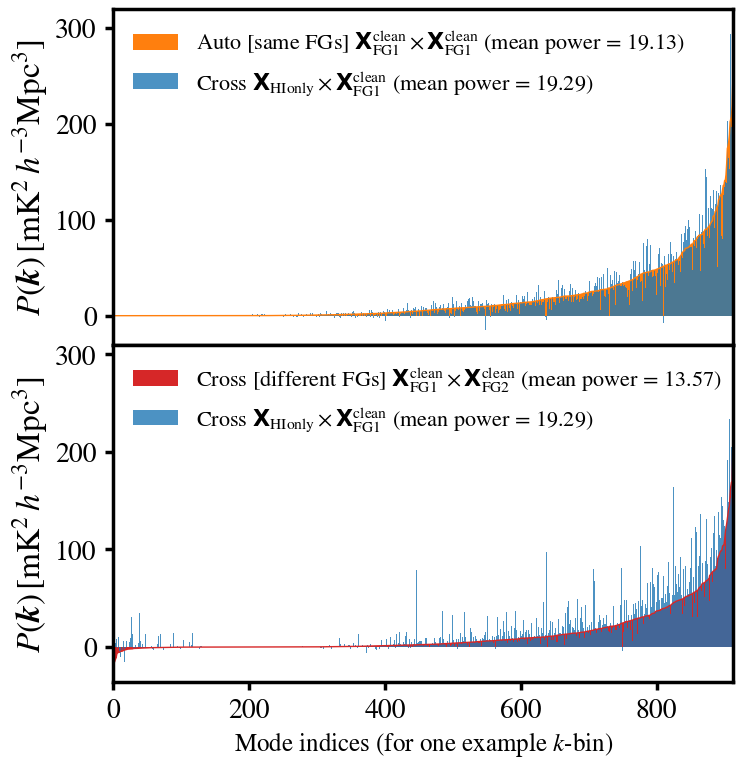}
    \caption{Amplitude of all Fourier modes $P(\boldsymbol{k})\txteq{\propto}\operatorname{Re}\left\{\tilde{\textbf{\textsf{X}}}_\text{A}(\boldsymbol{k}){\cdot}\tilde{\textbf{\textsf{X}}}_\text{B}^*(\boldsymbol{k})\right\}$ in the range $0.0590\txteq{<}k\txteq{<} 0.0687\,\hMpc$, which corresponds to one chosen $k$-bin in the spherically averaged power spectrum. The average of these modes, stated in the legend for each scenario, will represent the spherically averaged power spectrum value for the particular $k$-bin. Top-panel shows the equivalence in signal-loss between auto- and cross-correlation. Lower panel shows how signal loss can be larger where a different set of modes has been projected out in the foreground clean shown by the red results which is the cross-correlation between two different simulated foreground regions.}
    \label{fig:DiscreteSignalLoss}
\end{figure}

To provide a deeper understanding for why signal loss to the power spectrum is the same for auto- and cross-correlations, we present in \autoref{fig:DiscreteSignalLoss} the amplitude of all 3D-Fourier mode products for a randomly chosen $k$-bin. The spherically averaged $P(k)$ value for the chosen $0.0590\txteq{<}k\txteq{<} 0.0687\,\hMpc$ bin is then simply the average of all these amplitudes, which is stated in the legend for each scenario. The top-panel shows the comparison between an auto-correlation with foreground cleaning and the cross-correlation between foreground cleaned and foreground free (\hi-only) data. As can be seen, the average of the modes is approximately the same in both cases and thus consistent with \autoref{fig:AutoVsCross}, demonstrating that signal loss is equivalent in auto- and cross-correlation. The reason for this is related to the fact that the same modes are projected out of the analysis in both cases and signal loss does not compound when two maps with the same removed modes are combined in an auto-correlation. We confirm this to be the case in the bottom panel of \autoref{fig:DiscreteSignalLoss} where we \textcolor{black}{use} simulations with a foreground from a different region of sky so that we produce a cleaned map $\textbf{\textsf{X}}^\text{clean}_\text{FG2}$ which will have different foreground modes removed compared to the original $\textbf{\textsf{X}}^\text{clean}_\text{FG1}$ used in the rest of the paper for the MD1GPC simulation. The exact regions are not overly important just the fact that they will generate a different set of modes which are projected out in the foreground clean. When these two foreground cleaned maps are cross-correlated (red results) we now get a drop in power relative to the cross-correlation between $\textbf{\textsf{X}}^\text{clean}_\text{FG1}$ and the \hi-only map (see the mean power in the legend) showing that the signal loss is related to the modes being projected out in the foreground clean, and it is only a difference in these which will create further signal loss in a \hi\ auto-correlation.

The results demonstrated by \autoref{fig:DiscreteSignalLoss} have consequences for auto-correlation analyses with \hi\ intensity mapping. Not just because it further confirms that signal loss should be the same in auto- and cross-correlation where the same foreground modes are projected out, but also because the results in the bottom panel show when two differently cleaned maps are cross-correlated, the signal loss becomes more complex to estimate. This is relatable to a method that is likely to be pursued when attempting an auto-correlation detection whereby cross-correlations are measured between different sub-sets of the observations, created either by splitting data into different time-blocks ("sub-seasons") as done in GBT analysis \citep{Masui:2012zc,Wolz:2021ofa}, or by splitting different dishes as is possible with a multi-dish telescope such as MeerKAT. This is pursued in order to avoid the additive biases from noise and time- or dish-dependent systematics. Whilst this method would still observe the same foreground, the response to systematics may be different in each sub-set creating a scenario similar to the red results in \autoref{fig:DiscreteSignalLoss}. We leave further investigation into this specific form of auto-correlation method for future dedicated studies.  

\section{Applications to pathfinder intensity mapping with MeerKLASS}\label{sec:MK}

In the previous sections, the MD1GPC simulations used have been deliberately kept free of further observational effects (except for a couple of identified cases) besides the foreground contamination which included simulated polarisation leakage. However, a relevant question for current pathfinder single-dish experiments is whether the early pilot surveys, which typically have low signal-to-noise and additional systematic observational effects, can also rely on the foreground transfer function to correct for signal loss. In these pathfinder surveys 
\cite[e.g.][]{Wolz:2021ofa,Cunnington:2022uzo} foreground cleaning is typically aggressive and signal loss can reach high levels, thus one could argue that we are more reliant on signal reconstruction in these early surveys, compared with future surveys where foreground cleaning and systematics will be more controlled.

The cosmological detection in \citet{Cunnington:2022uzo} (like all other intensity mapping detections preceding it) relied on a foreground transfer function to reconstruct the signal loss from foreground cleaning. The $7.7\sigma$ cross-correlation detection significance fell to ${\sim}\,4\sigma$ without signal reconstruction. The survey covered just $200\,\text{deg}^2$ and only the frequency channels spanning $1015-973\,\text{MHz}$ (0.4\txteq{<}z\txteq{<}0.46) were used to avoid the worst RFI. Furthermore, the observation gathered just $10.5\,\text{hours}$ of data per dish. This means sky coverage and signal-to-noise were low and foregrounds could be easily impacted by systematics, rendering signal loss more complex and widespread than the examples we have investigated so far. 

The reliability of the transfer function was validated for the results in \citet{Cunnington:2022uzo} and here we demonstrate these validation tests by utilising a different set of simulations which we refer to as the MDMK simulations, which aim to emulate the MeerKAT 2019 pilot intensity mapping survey \citep{Wang:2020lkn}. We use the survey's non-uniform mask and use fluctuations in the uncleaned foreground sky to generate systematic perturbations in the MDMK foreground simulations creating a demanding cleaning requirement. We outline the details of how this is achieved in the following section.

\subsection{MDMK MeerKLASS simulations}\label{sec:MDMK}

To emulate current MeerKAT pathfinder data and investigate the performance of the transfer function when signal loss is spread across a wide range of scales, we utilise the MeerKAT 2019 pilot survey data \citep{Wang:2020lkn,Cunnington:2022uzo}. The survey targeted a single patch of ${\sim}\,200\,\text{deg}^2$ in the WiggleZ 11hr field, covering $153^\circ\,{<}\,\text{R.A.}\,{<}\,172^\circ$ and $-1^\circ\,{<}\,\text{Dec.}\,{<}\,8^\circ$. The telescope observed at constant elevation, scanning back and forth through azimuth taking $1.5\,\text{hours}$ to complete one time-block. 7 time-blocks were obtained with a mix of rising and setting scans, creating offset coverage providing the footprint which can be seen in the MDMK simulation maps in \autoref{fig:MKmaps}. Holes in the footprint are evident and are caused from the multi-stage RFI flagging which can leave gaps in the scanning. For the MeerKAT \hi\ simulation, (top-panel of \autoref{fig:MKmaps}), we use the same \textsc{Multi-Dark} simulation as in MK1GPC (\secref{sec:MD1GPC}) but calculate the physical volume covered by the MeerKAT 2019 data and cut a volume of this size from the $1\,(\text{Gpc}/h)^3$ cube. We use a similar pixelisation as the 2019 data ($n_\text{x},n_\text{y},n_\text{z}\txteq{=}133, 41, 250$), and then apply the exact same footprint mask. The MeerKAT 2019 observations were performed in L-band but to avoid dominant RFI, only 199 channels with a $973.2-1014.6\,\text{MHz}$ frequency range ($0.400\,{<}\,z\,{<}\,0.459$) were used. Additionally, of the 199 channels selected to use, a further 32 are removed due to evidence of RFI contribution in their eigenmodes. We also replicated this exact channel flagging in the MDMK simulations. 

\subsubsection{Frequency perturbed foregrounds}\label{sec:PerturbedFGs}

Evidence of systematics was seen in the MeerKAT 2019 data, which was expected given the low amount of observational time and it being a first of its kind pilot survey. One way systematics were evident was in the perturbations to what should be smooth spectra in the raw foreground sky. The exact cause of these systematic perturbations is beyond the aim of this paper but we can still use the distorted spectra from the real data to perturb an idealised foreground simulation, emulating the main impact from these systematics on the foreground clean and signal reconstruction.

\begin{figure}
    \centering    \includegraphics[width=1\linewidth]{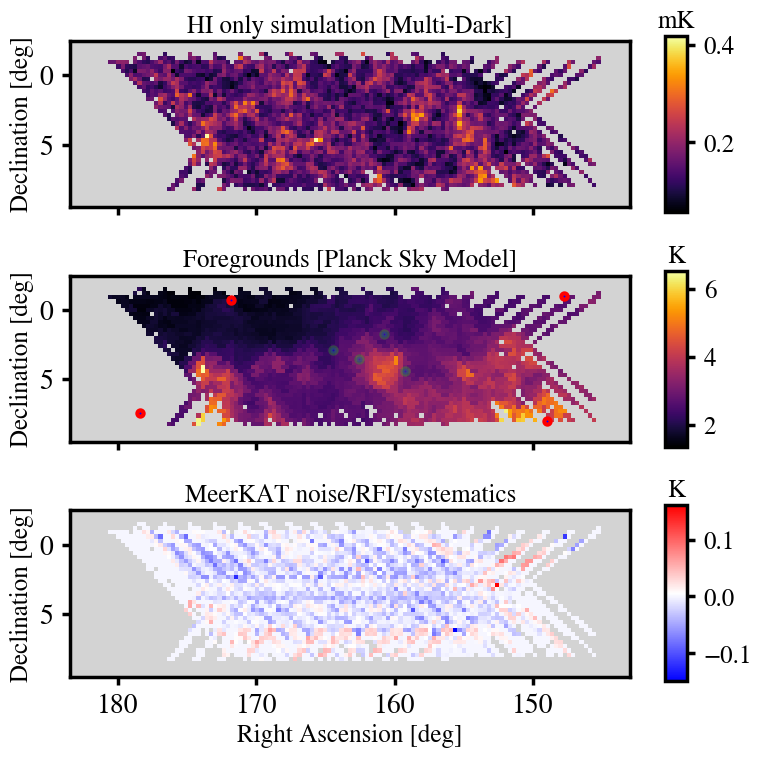}
    \caption{MDMK simulated maps aiming to emulate MeerKAT 2019 intensity mapping data, averaged along the $973.2{-}1014.6\,\text{MHz}$ frequency range. Top panel shows HI only, produced using the \textsc{Multi-Dark} $N$-body semi-analytical simulation. Middle-panel shows the foregrounds simulated using a perturbed version from the Planck Sky Model. The points on the foreground map indicate the positions of the example spectra plotted in \autoref{fig:MKspectra}, with green and red points representing pixels near the centre or edge of the MeerKAT footprint respectively. The bottom panel shows an estimate for some MeerKAT noise and residual systematics obtained by subtracting data observed at different times (see \secref{sec:MKnoise}).}
    \label{fig:MKmaps}   \includegraphics[width=1\linewidth]{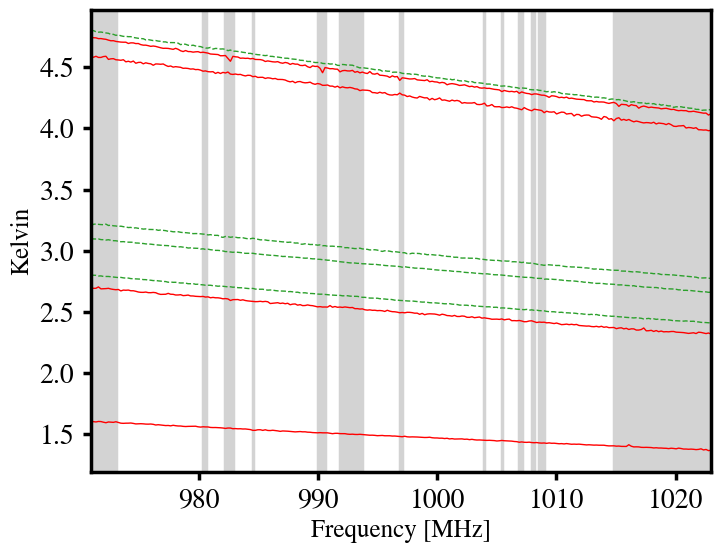}
    \caption{Example spectra from the MDMK foreground simulations aiming to emulate MeerKAT pilot survey data. The perturbations to the spectra have been produced using the MeerKAT 2019 data to create realistic foreground simulations with systematic effects (see \secref{sec:PerturbedFGs}). The green-dashed lines represent pixels taken from a central position in the MeerKAT footprint (corresponding to green points in \autoref{fig:MKmaps}). The red-solid lines represent pixels near from the edge of the footprint (corresponding to red points in \autoref{fig:MKmaps}) and are more vulnerable to systematics, hence the more noticeable perturbations. The grey-shaded regions represent channels which were flagged in the cross-correlation analysis \citep{Cunnington:2022uzo}, which we also flag in this simulation for consistency.}
    \label{fig:MKspectra}
\end{figure}

To create foregrounds for the MDMK simulations we begin by using the Planck Sky Model to generate synchrotron and free-free emission at the relevant frequencies and sky position, as with the MD1GPC simulation. The bottom panel of \autoref{fig:MKmaps} shows the foreground simulation for the MeerKAT 2019 footprint. Unlike the MD1GPC simulation, we do not include any simulated polarisation leakage and instead use the MeerKAT 2019 data itself, aiming to create more realistic systematic perturbations to the foregrounds. This is done by fitting a smooth polynomial to each line-of-sight in the 2019 data, then the systematic perturbations to the foreground spectra can be approximated by the ratio between the data and the polynomial i.e.
\begin{equation}\label{eq:PerturbedMKSims}
    T_\text{perturbation}(\boldsymbol{x},\nu) = \frac{T_\text{2019-data}(\boldsymbol{x},\nu)}{T_\text{smooth-poly}(\boldsymbol{x},\nu)}\,.
\end{equation}
We found on average these perturbations were small sub-percent values however they could be as high as $3.7\%$. We multiply these perturbations with the PSM foreground and \autoref{fig:MKspectra} shows some example perturbed spectra from the final simulation. The perturbations are worse near the edge of the map (shown as red-solid lines) where due to the lower coverage, systematics can have more impact. This is consistent with what was found in the actual data and in the cross-correlation analysis these edge pixels are down-weighted \citep{Cunnington:2022uzo}. \autoref{fig:MKspectra} also shows the flagged channels (grey regions) used in the cross-correlation analysis which we also adopt in the MDMK simulation.

\subsubsection{Anisotropic systematics and RFI residuals}\label{sec:MKnoise}

The analysis of the auto-correlation power spectrum for the 2019 MeerKAT survey in \citet{Cunnington:2022uzo} showed evidence of additive biases most likely from instrumental noise, residual RFI, or other systematics. Their contribution appears to dominate over the \hi\ signal because the auto-power spectra amplitude was larger than one would expect from \hi\ only power. We also include a contribution to the MDMK simulation which attempts to emulate these types of additional components. Again, we use the real MeerKAT 2019 survey itself to produce a map of time-varying anisotropic contributions and add these onto the MDMK \hi\ and perturbed foreground maps. This is achieved by taking the residuals from different time blocks in the MeerKAT 2019 survey. We take the difference between the first four time-blocks and the last three where these residuals will represent components that vary in time. Therefore, in principle, this should not include the \hi\ signal or the foregrounds since these would be consistent in time, but instead only include time-varying systematic contributions, which is what we are aiming to emulate.

The map of the MeerKAT time-block residuals is shown in the bottom panel of \autoref{fig:MKmaps}. In some instances, there are no shared pixels in both time-block groups so the residual is undefined. For these pixels we instead resort to adding a large level of Gaussian random noise whose variance dominates the \hi\ signal by one order of magnitude. However, when these are plotted in \autoref{fig:MKmaps}, which is the average along the line-of-sight, their contribution is averaged down which is why the amplitude appears relatively low around the edges where most of the missing pixels between time-blocks are. The pixels from the actual MeerKAT residuals, which are most concentrated in the centre where shared coverage is better, appear higher relative to the Gaussian noise. This will be due to the residuals being more correlated in frequency, thus do not average down in the plot. The frequency correlation and apparent anisotropies of the residuals as evident in \autoref{fig:MKmaps}, suggest they are contributions beyond simple instrumental noise. Whilst this is a complication for the pilot survey analysis, it is useful for our purposes, providing additional complications to foreground cleaning and signal loss in the MDMK simulations.

In these simulations, whilst we do not explicitly include the effects from a realistic telescope beam, some of the impacts it has on the foregrounds will be included in the perturbations we add to the simulations from \autoref{eq:PerturbedMKSims}. A simple Gaussian beam is trivial to include and assuming it is approximately matched in the transfer function construction, it makes no difference to the performance of the transfer function as we will explicitly show in the following section \secref{sec:ParameterDep}. However, in reality, the MeerKAT beam will be more complex with wide-reaching frequency-dependent side lobes which could complicate foreground cleaning \citep{Matshawule:2020fjz,Spinelli:2021emp}. Trying to replicate this in the mocks in the transfer function construction may be difficult and an investigation into what level this needs to be considered should be pursued. This requires a more detailed simulation which we will pursue in follow-up work.

\subsection{Correcting signal loss in pathfinder data}

\begin{figure}
    \centering
    \includegraphics[width=1\linewidth]{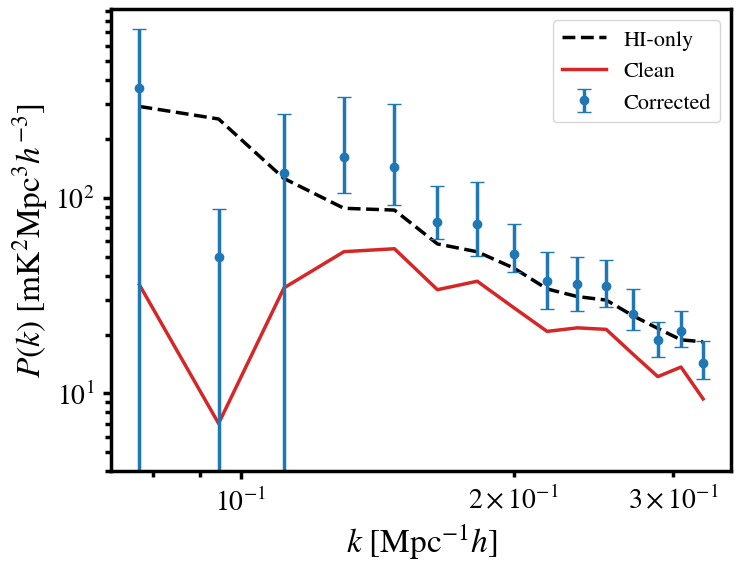}
    \caption{Power spectra for MDMK simulations which are small in volume, have a more complex non-continuous survey footprint, high anisotropic noise and residual systematics, as well as systematically perturbed foregrounds, all to emulate actual MeerKAT pathfinder data. We show the cross-correlation with the \hi-only maps. Signal loss from the $\Nfg\txteq{=}10$ PCA clean (red line) is larger and more widespread into high-$k$ compared with previous results in the more idealised MD1GPC simulation. Despite this the transfer function is encouragingly still able to reconstruct a reasonably unbiased result shown by the blue data points. Error bars are given by the limits of the central 68th percentile region from the distribution of reconstructed power spectra using the transfer function mocks.}
    \label{fig:MKpower}
\end{figure}

\autoref{fig:MKpower} shows power spectra for the MDMK (MeerKAT-based) simulations presented in the previous sub-section. The black dashed line shows the foreground-free \hi-only result, and the red solid line shows the result from adding the perturbed foregrounds and residual time-varying systematics based on real MeerKAT data, then performing a $\Nfg\txteq{=}10$ PCA clean. We find that using the perturbed foregrounds (described in \secref{sec:PerturbedFGs}) is the main cause for requiring an aggressive foreground clean, highlighting the importance of instrument calibration so that smooth spectra are maintained in the observed data. Similar to MeerKAT data \citep{Cunnington:2022uzo}, signal loss in the foreground cleaned data is widespread throughout all scales, with noticeable signal loss occurring even in the highest-$k$. The main reason for this is due to the decreased depth of the frequency/redshift range. We tested this with the main MD1GPC simulation. Reducing the number of pixels along the LoS by a factor of 4 to 64 pixels with $0.25\,\text{Gpc}/h$ of depth,  produced ${>}\,75\%$ signal loss in the smallest 4 $k$-bins, and still $13\%$ in the highest $k$-bins. This was even without the polarisation leakage and just removing 4 PCA modes. Whereas for an equivalent scenario but using the full $1\,\text{Gpc}/h$ depth, the signal loss is never greater than $30\%$ and only $3\%$ at the highest-$k$. 
This can be understood by considering that modes projected out of narrow frequency-range data will be confined to a higher-$k_\parallel$ space. Thus the signal loss will also spread into higher $k$-modes. Encouragingly this means that signal loss should be naturally mitigated in future observations by using a larger frequency range. This will be possible with MeerKAT UHF-band observations which will probe lower frequencies where RFI is expected to be less dominant, thus a more complete frequency range can be used.

Despite the more complex and widespread signal loss in \autoref{fig:MKpower} (red line), when we construct a transfer function using the process summarised in \secref{sec:TFsummary}, we are able to reconstruct the correct \hi\ power spectrum. As in previous tests, the clean and corrected power spectra are cross-correlations with the original-\hi\ to avoid any issues with residual foreground contamination confusing the assessment of signal loss. The power spectra are naturally more noisy than the previous MD1GPC simulations due to the decreased volume, the systematically perturbed foregrounds (see \autoref{fig:MKspectra}) and the large time-varying systematics (\autoref{fig:MKmaps} bottom panel) inserted into the simulation. The instrumental noise and additive systematics is an important consideration that we did not include in the default MD1GPC simulations of previous sections. This additional noise will introduce extra perturbations to the foreground modes, in the same way the signal introduces perturbations (see \autoref{fig:PerturbedEigenvecs}). If the noise is large, as is the case for pathfinder observations with low observational time, these perturbations will be large. Encouragingly, this does not appear to cause noticeable problems for the transfer function, evidenced by the corrected result in \autoref{fig:MKpower}, which includes large additional contributions that dominate over the \hi\ signal.

Given the more complex nature of the pilot survey simulation, some reconstructed power spectra from the transfer function mocks produced outliers and returned non-Gaussian distributions for each $k$-mode. In this scenario, using the rms over the mocks for the errors would be a poor estimation and would be overly distorted by the outliers. We therefore instead use the 68th percentiles limits to provide the asymmetric error bars, which is what are presented in \autoref{fig:MKpower}. To obtain the converged distribution, we used 1000 mocks in the transfer function calculation. This presents a further advantage of using the transfer function mocks for error estimation, providing more options to handle non-Gaussian uncertainties. This of course would have complex implications for further analysis and parameter estimation, which we do not investigate here. However, errors should naturally become more Gaussian as intensity map quality improves and noise, systematics, etc. are reduced.

\begin{figure}
    \centering
    \includegraphics[width=0.9\linewidth]{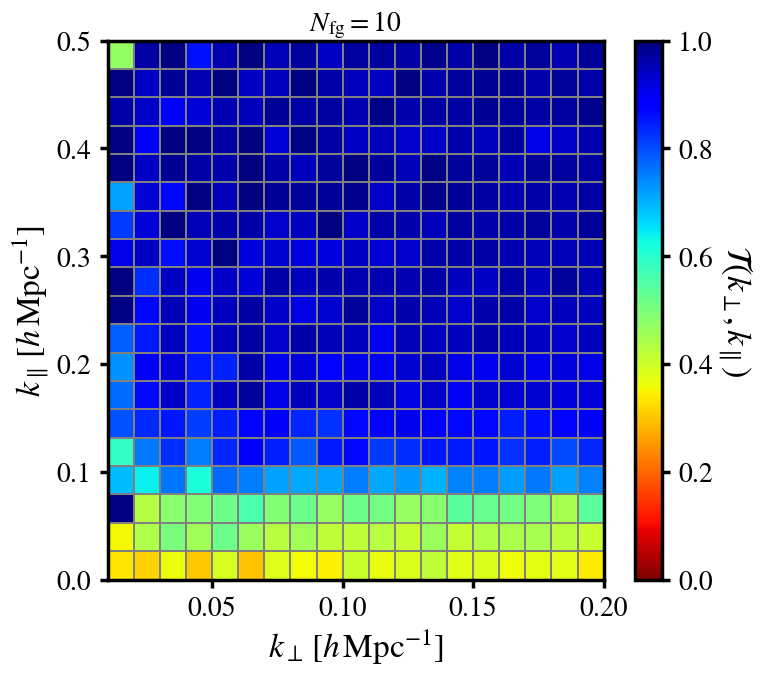}
    \caption{Foreground transfer function in cylindrical $k_\perp,k_\parallel$ space for the MDMK simulations emulating MeerKAT pathfinder data. $\Nfg\txteq{=}10$ PCA modes have been removed for the foreground clean. The 1D spherically averaged version of this is used for the corrected results in \autoref{fig:MKpower}.}
    \label{fig:MK2DTF}
\end{figure}

\autoref{fig:MK2DTF} shows the computed transfer function in cylindrical $k_\perp,k_\parallel$ space for the MDMK simulations. It is interesting to analyse the differences between this more realistic case and that from the more idealised MD1GPC simulations in \autoref{fig:TFdemo2D}. This again reveals that signal loss is more widespread into larger $k_\parallel$ modes, which is consistent with the more widespread signal loss evident in \autoref{fig:MKpower}. We still see large regions where $\mathcal{T}\txteq{\sim}1$ suggesting that the approach of discarding some regions of $k$-space to massively reduce the dependence on the transfer function (as we discussed in \secref{sec:TF}) could still be pursued even with small intensity mapping pilot surveys.

The results from the MDMK simulations provide validation for using the foreground transfer function with pathfinder survey intensity maps. We have used MeerKAT data itself to attempt to complicate the foreground clean and signal loss to stress-test the current signal reconstruction process. The simulations produced make no attempt to understand the source of the perturbations to foreground spectra or additive time-varying systematics, simply emulating them as realistically as possible to mimic the challenge they pose. Whilst the success is encouraging, the investigation would be completed further by including specific simulations of known observational effects such as directly simulating RFI, 1/$f$ correlated noise, non-uniform noise, a non-Gaussian beam etc. This would allow more analysis into exactly what observational effects are most troublesome for foreground cleaning and signal loss. The development of a robust simulation pipeline for MeerKAT single-dish intensity maps including a realistic beam and all these observational effects is outside the aims of this paper but is being pursued in other MeerKLASS collaboration projects \citep[e.g.][]{Irfan:2023njr}.

\section{Precision cosmology suitability}\label{sec:cosmo}

In this section we look to future \hi\ intensity mapping observations and test how reliable a transfer function would be where sub-percent accuracy on parameter estimates is required
. We return to the more generic simulations of MD1GPC to avoid the investigation being confused by the large statistical noise present in the previous section's realistic simulations of a MeerKAT pilot survey.

\subsection{Mock parameter dependence}\label{sec:ParameterDep}

Up until now, it has not been investigated how robust the accuracy of the transfer function is when there are discrepancies between parameters used in the transfer function construction and their true fiducial values in the real observed data. If the parameter assumptions used for the generation of 100 mocks strongly influence the final accuracy of the reconstructed power spectrum then this is a large concern for precision cosmology since this would lead to biased cosmological parameter estimates. 

In this section, we demonstrate how mild the mock parameter dependency is and show how large $+/-\,100\%$ discrepancies between the assumed parameters in the mocks used for transfer construction and the underlying truth in the data, mostly only yield small ${\lesssim}\,1\%$ inaccuracies in the recovered parameter estimations. We test this by treating the MD1GPC as the observed data with the underlying "true" parameters, then vary some of the values $\{\Omega_\hi\txteq{\propto}\overline{T}_\hi,f,\sigma_\text{v},R_\text{beam}\}$ in the lognormal mocks which are used to construct the transfer functions. The model power spectrum we use to generate the lognormal mocks is described in more detail in \appref{sec:HIPkMod} but we repeat it here for convenience
\begin{multline}\label{eq:Pkmodelmaintext}
    P_\text{mod}(k,\mu) = \overline{T}^2_\hi \frac{\left(b^2_\hi+f\mu^2\right)^2}{1+(k\mu\sigma_\text{v}/H_0)^2} P_\text{m}(k)\\
    \times\,\exp\left[-(1-\mu^2)k^2 R_\text{beam}^2\right]\,.
\end{multline}
For this test we used the MD1GPC simulations with a $\sigma_\text{n}\txteq{=}1\,\text{mK}$ dominant white noise and a default Gaussian beam where $R_\text{beam}\txteq{=}10\,h^{-1}\text{Mpc}$.

\begin{figure}
    \centering
    \includegraphics[width=1\linewidth]{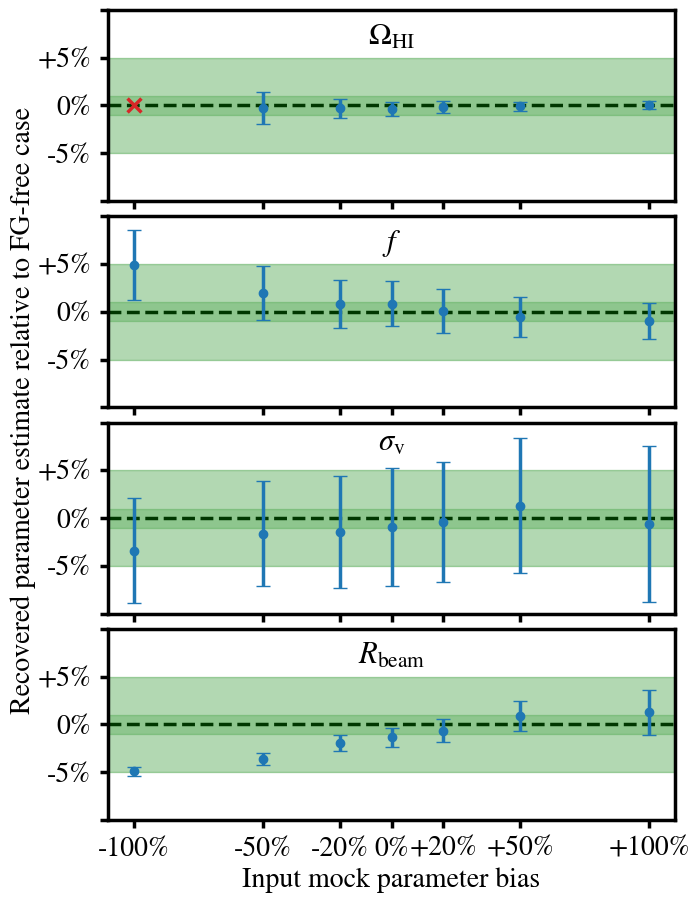}
    \caption{Robustness of the transfer function in response to biased parameter assumptions. Each panel shows a different parameter used in the mocks which construct the transfer function. The $x$-axis shows different \%-biases relative to the correct fiducial values. The $y$-axis shows the estimated parameter posterior from an MCMC on the reconstructed data using a transfer function. The error bars represent the 68\% confidence region in the estimated posteriors. These are plotted relative to the foreground-free parameter estimates to demonstrate that only small biases are induced from incorrect parameter assumptions in the transfer function construction. These results are produced with the MD1GPC simulation with $\sigma_\text{n}\txteq{=}1\,\text{mK}$ dominant noise and a Gaussian beam with $R_\text{beam}\txteq{=}10\,h^{-1}\text{Mpc}$. All results are for a $\Nfg\txteq{=}12$ PCA clean.}
    \label{fig:ParameterResponse}
\end{figure}

\autoref{fig:ParameterResponse} shows how subtle the impact on parameter estimation is when parameters used in the mock's model power spectra, given by the panel titles, are biased by an amount indicated by the $x$-axis. The $y$-axis shows the percentage bias relative to the foreground-free parameter estimation. In all cases, we sample the parameter posterior distribution in a Bayesian MCMC only varying one parameter at a time, fixing all other parameters in the model (\autoref{eq:Pkmodelmaintext}) to fiducial values fitted to the foreground-free MD1GPC simulation. The error bars represent the 68\% confidence regions in the posterior distributions and they are plotted relative to the median of the foreground-free posterior. To avoid non-linear complications in the modeling we only use modes where $k\txteq{<}0.3\,\hMpc$. Whilst these scales are still quite non-linear, our model worked reasonably well on these scales and since we are testing its performance relative to the foreground-free case, any shortcomings caused by non-linear effects will be present in both. 

$\Omega_\hi$, which is proportional to $\overline{T}_\hi$ (see \autoref{eq:TbarModelEq}), will only change the amplitude of the \hi\ power spectrum in our simple linear model, and the transfer function appears extremely robust to these scale-independent amplitude changes, with no ${>}\,1\%$ bias being induced. We even tested going to $4\times$ the fiducial truth on $\Omega_\hi$, since this is a very unconstrained parameter, but this still yielded a sub-percent bias. Note that the red-cross indicates an undefined result for the $\Omega_\hi\txteq{-}100\%$ case since the intensity maps are zero when $\overline{T}_\hi\txteq{\propto}\Omega_\hi\txteq{=}0$. The increase in uncertainties with a decreasing $\Omega_\hi$ mock parameter input is caused by the white noise having a more dominant impact relative to these lower amplitude mocks in the transfer function construction.

The growth rate $f$ is included to model the redshift-space distortions (RSD) which are present in the MD1GPC simulations. $f$ will be an anisotropic parameter since RSD are a line-of-sight-only effect. This appeared to induce more of a bias in the reconstructed power spectra, but it still only caused ${<}\,2\%$ in most cases tested. For certain cosmological parameters such as the growth rate, much tighter priors will be applicable such that the biased parameter values we have used for $f$ will be unrealistic. It is interesting to see that failing to include a linear RSD model in the mocks at all, as shown by the ${-}100\%$ result (i.e. $f\txteq{=}0$), induces a ${\sim}\,5\%$ bias which suggests that it is important to include some basic anisotropic RSD in the mocks for reconstruction accuracy.

$\sigma_\text{v}$ is again used to model RSD but as a phenomenological attempt to model fingers-of-god (FoG) on mildly non-linear scales. This will also be an anisotropic parameter but is also directly scale-dependent too, having a greater influence at high-$k$. This shows ${\sim}\,3\%$ bias if the parameter is set to zero. Even though this is a phenomenological parameter without a physically defined fiducial value, it is still unlikely that the parameter will be completely omitted without a suitable method for replacing its modelling effects. We highlight that the increase in error bars for the $f$ and $\sigma_\text{v}$ parameters is not necessarily foreground-related. Some parameters will be constrained better than others and $\sigma_\text{v}$ relies more on small scales which are damped by the beam. Furthermore, both parameters would be more suitably constrained by including the quadrupole as opposed to the spherically averaged monopole we are using here. 

Lastly, we introduce the size of the beam $R_\text{beam}$ as a varying nuisance parameter. This is defined as the rms of the beam profile in comoving units at the probed redshift. Similarly to $\sigma_\text{v}$ this is another scale-dependent anisotropic parameter, although this time affecting high-$k_\perp$ modes. This can reach a ${\sim}\,5\%$ negative bias if completely unaccounted for, shown by the ${-}100\%$ result. For nuisance parameters linked to the instrument such as $R_\text{beam}$, we should have a much tighter prior on its value effectively ruling out a ${>}20\%$ incorrect assumption. The small ${\sim}\,1\%$ bias relative to the truth in the case where the correct fiducial beam has been used (see $0\%$ input mock parameter bias result), suggests there could be some discrepancy in high-$k_\perp$ modes between foreground-free and reconstructed foreground-cleaned data, where $R_\text{beam}$ has the most impact, leading to inaccuracies in its estimation. Given this is only a very small bias and is only in a nuisance parameter, we defer this to future work, where we will investigate the impact on signal reconstruction in the presence of more complex beams. 

Given that the small biases appear to be caused by anisotropic divergences between mocks and observations, we investigated whether the 2D transfer function (discussed in \autoref{fig:AnisotropicTF}) could yield improvements. We found that when there is no polarisation leakage in the simulations, the average bias on the reconstructed power spectrum from using $f\txteq{=}0$ in the mocks is $3.8\%$ relative to the foreground-free power spectrum. Interestingly though, when we construct and apply the transfer function in 2D, then rebin into 1D following the same procedure in \autoref{eq:2Dto1D}, the bias is only $1.9\%$. However, as we found from the results in \autoref{fig:AnisotropicTF}, the variance blows up when we reintroduce the more complex foreground with polarisation leakage. We thus leave further investigation to future work with more realistic simulations where we will definitely test if the large variance in the 2D transfer function can be reliably reduced, and if it then still decreases the small biases from anisotropic inconsistencies we see in \autoref{fig:ParameterResponse}.

There are some important conclusions to draw from the results in \autoref{fig:ParameterResponse}. Firstly, the results demonstrate how the transfer function works. It is not a process where an exact replica of the real data is required to measure the precise impact of signal loss. Rather the mocks injected act as a test field to construct a response function caused by the foreground cleaning. Broadly speaking, it appears that the parameters used in the mocks for the transfer function do not have a strong influence on the final accuracy of the reconstructed power spectrum. However, where sub-percent accuracy is the aim, it is clearly beneficial to have the mocks attempt emulation of some of the features inherent in the observational data. For example, completely neglecting the telescope beam or linear RSD in the mocks, can have a noticeable impact on the reconstruction accuracy. This lays the foundation for many further inquiries into this topic. For example; will a more realistic frequency-dependent beam with a side-lobe structure be sufficiently emulated by a Gaussian beam in the mocks for the purposes of the foreground transfer function? Do any of the other multitude of parameters that we want to probe or are forced to include in our model as nuisance parameters, have a stronger influence on reconstruction accuracy? What happens when we allow multiple parameters to vary simultaneously as apposed to varying one parameter at a time as done for the results in \autoref{fig:ParameterResponse}? Whilst these results are strong evidence that a transfer function will not bias parameter inference, the most robust way to confirm this would be with more realistic simulated observational effects e.g. a MeerKAT model of the beam, and to perform detailed modelling with a multi-parameter MCMC fit to the reconstructed power spectrum, including \textit{shape} parameteres e.g. $h$, $n_\text{s}$, $\omega_\text{c}$, $\omega_\text{b}$, under a range of different scenarios. This is beyond the scope of this work but is something we will aim to showcase in a follow-up study.

Since it seems the fiducial mock parameters have a sub-dominant influence over the reconstruction accuracy, one prospect to consider is a process whereby the mock parameters are updated based on the parameter inference from the real data. In this way an iterative transfer function could be developed whereby as the parameter posteriors for the observed data are estimated and converge on a final parameter estimate, these values can be used to update the transfer function calculation and avoid any possibility of it biasing the parameter inference. We discuss some of the early investigations for this possibility in \appref{sec:MCMCTF}, but also largely leave this to further dedicated investigation.

\subsection{Probing exotic physics on ultra-large scales}\label{sec:fNL}

As a final test of the transfer function's performance, we examine its ability to reconstruct signal on the largest scales, even when the underlying true signal has some unknown "non-standard" properties. For this test, we focus on primordial non-Gaussianity (PNG) which can be probed on the largest scales in galaxy surveys \citep{Mueller:2022dgf} and soon in \hi\ intensity maps \citep{Li:2017jnt,Witzemann:2018cdx,Karagiannis:2020dpq}.

The nature of the fluctuations in the primordial Universe which arise during inflation carry a wealth of information regarding the physical mechanisms that shaped the early Universe. The parameter $f_\text{NL}$ quantifies the departure from Gaussianity in the primordial Universe \citep{Komatsu:2001rj} and for the so-called local-type of PNG, $f_\text{NL}\txteq{\neq}0$ would be evidence of non-Gaussian fluctuations, ruling out slow roll, single-field inflation in favour of more exotic multi-field models \citep{Creminelli:2004yq}. Constraints on PNG so far come from CMB anisotropies and results are consistent with Gaussian fluctuations with $f_\text{NL}\txteq{=}0.9\txteq{\pm}5.1$ \citep{Aghanim:2018eyx}. However, large-scale structure surveys, in particular intensity mapping, are expected to soon lead the way in improving PNG precision. Evidence for PNG in large-scale structure surveys will manifest as a scale-dependent correction to the linear bias \citep{Dalal:2007cu}. This correction scales as $k^{-2}$ thus it is at \textit{ultra}-large scales where sensitivity to $f_\text{NL}$ becomes most prominent.

In this section, we generate a new underlying \hi\ intensity map simulation, no longer using the MD1GPC simulation. The reason for this is first because we wish to add a clear signature of PNG into the field, and secondly, because we need to cover much larger scales where we will be able to probe the scales that are sensitive to the $f_\text{NL}$ parameter. We use the $N$-body COmoving Lagrangian Acceleration (COLA)\footnote{\href{https://pypi.org/project/pycola3/}{pypi.org/project/pycola3}} \citep{Tassev:2013pn,Tassev:2015mia}, code to generate a fast $N$-body simulation on a $(8{,}000\,h^{-1}\text{Mpc})^3$ grid with $256^3$ pixels which approximately represents a wide and deep \hi\ intensity mapping survey with something like SKAO. We seed the simulation with a \hi\ power spectrum given by 
\begin{equation}\label{eq:fNLPk}
\begin{aligned}
    P_\hi(k,&\mu,z) = \\
    &\overline{T}_\hi^2(z)\left[b_\hi(z)+\Delta b_\hi(k,z) f_{\mathrm{NL}}+f(z) \mu^2\right]^2 P_{\mathrm{m}}(k, z)\,,
\end{aligned}
\end{equation}
where
\begin{equation}
    \Delta b_\hi(k,z) = \left[b_\hi(z)-1\right] \frac{3 \Omega_{\mathrm{m}} H_0^2 \delta_{\mathrm{c}}}{c^2 k^2 T(k) D(z)}
\end{equation}
and $\delta_\text{c}\txteq{=}1.686$ is the critical matter density contrast for
spherical collapse, $T(k)$ is the \textit{matter} 
(not foreground) transfer function, and lastly the growth function can be defined by
\begin{equation}
    D(z)=\frac{5}{2} \Omega_{\mathrm{m}} H_0^2 H(z) \int_z^{\infty} \frac{1+z^{\prime}}{H^3\left(z^{\prime}\right)} \mathrm{d} z^{\prime}\,.
\end{equation}
We set $f_\text{NL}\txteq{=}100$ in \autoref{eq:fNLPk} to provide a clear scale-dependent bias on large scales in the new observed data simulation. We add foreground contamination to this following the same steps as the MD1GPC foreground model, run the PCA foreground clean, then construct the foreground transfer function following the same methods in the rest of this section. The lognormal mocks which construct the transfer function will assume $f_\text{NL}\txteq{=}0$ and we will be testing if this transfer function can still recover the true underlying \hi\ power spectrum with a scale-dependent bias induced by the $f_\text{NL}\txteq{=}100$ PNG signature. Despite  $f_\text{NL}\txteq{=}100$ being confidently ruled out by Planck18 observations, we still use this to set up a highly diverged underlying cosmology from that assumed in the transfer function construction, similar to the extreme biases we tested in \autoref{fig:ParameterResponse}.

\begin{figure}
    \centering
    \includegraphics[width=1\linewidth]{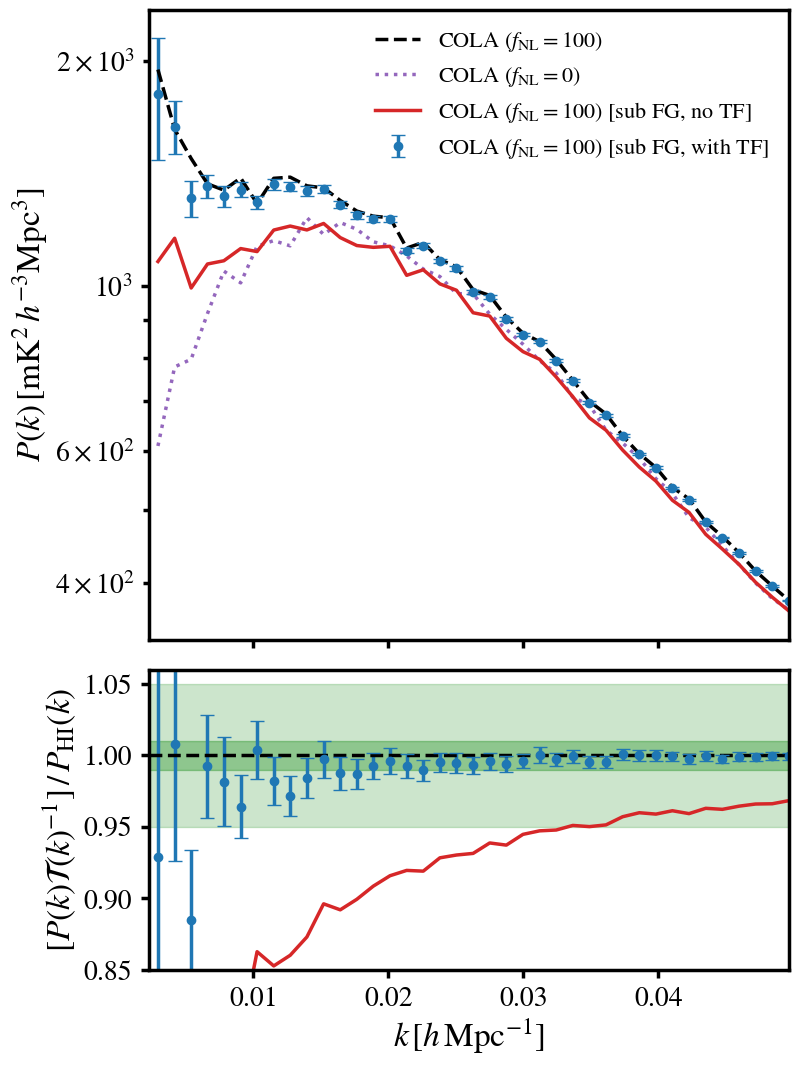}
    \caption{Capability to detect "non-standard" cosmology on ultra-large scales when the fiducial cosmology used in the transfer function construction assumed standard. For this test, we use the PNG parameter $f_\text{NL}$. We let the underlying truth have an extreme $f_\text{NL}\txteq{=}100$, produced using COLA simulations (black-dashed line). Despite an incorrect $f_\text{NL}\txteq{=}0$ assumption used in the mocks for the transfer function, the correct $f_\text{NL}$ signature is still recovered in the reconstructed power spectrum (blue data points) across most scales.}
    \label{fig:fNL}
\end{figure}

\autoref{fig:fNL} shows the results for the large-scale PNG COLA simulation. The black-dashed line shows the foreground-free result where the impact from the $f_\text{NL}\txteq{=}100$ input is clearly evident on small-$k$. For reference we also plot the $f_\text{NL}\txteq{=}0$ equivalent case shown by the purple-dotted line. Adding foreground contamination and cleaning is enough to greatly distort the PNG signature (shown by the red-solid line). For this simulation, we do not add polarised foregrounds, therefore only a $\Nfg\txteq{=}4$ PCA clean is required. This is more representative of a future large sky survey where it is safer to assume an enhanced level of calibration has been achieved. The blue data points show the result where we have used the transfer function to reconstruct the signal loss from the foreground clean. The errors are estimated from the variance of the transfer function as outlined in \secref{sec:ErrorEst}. There is good agreement between the reconstructed power spectrum and the true underlying power spectrum (black-dashed) where $f_\text{NL}\txteq{=}100$, shown in more detail by the residuals in the bottom panel. We emphasise that this is an extreme case where the underlying true cosmology is very different from that assumed by the mocks in the transfer function, and despite this only three of the measured modes in the reconstructed power spectrum are more than $1\sigma$ beyond sub-percent accuracy. 

Previous work in \citet{Cunnington:2020wdu} investigated signal loss from foreground cleaning in the context of PNG and demonstrated how a phenomenological model could be implemented to account for the signal loss. Whilst this delivered successful simulation-based results, there was a worrying degeneracy between the parameters in the signal loss model and $f_\text{NL}$, thus tight priors would be needed for such a method to allow good constraints on $f_\text{NL}$. These tight priors would only come from a very good understanding of foreground contamination and signal loss. The advantage behind the foreground transfer function approach is that no phenomenological model is required and the signal loss is reconstructed using the observed data itself without the need for additional nuisance parameters. We emphasise that this is a preliminary investigation into PNG with a foreground transfer function and a more complete study with robust large-scale simulations is required, including more detailed modelling and MCMC parameter estimation, which we defer to future work.

The previous results in \autoref{fig:ParameterResponse} supported the claim that a discrepancy between parameters assumed for the transfer function and the underlying truth in the data, only has a mild influence on results. \autoref{fig:fNL} extends this conclusion to the largest scales where foreground contamination is most troublesome and thus is a very encouraging result. A slight negative bias in the reconstructed power spectrum is apparent, likely caused by the lower $f_\text{NL}$ in the mocks, thus we cannot claim complete independence from the mocks. Sensible assumptions should therefore still be made when fixing a fiducial cosmology for the transfer function construction. However, there is potential for this small bias to be eradicated by adopting the iterative approach to transfer function calculation as discussed in \appref{sec:MCMCTF}. In this case, the recovered $f_\text{NL}$ inferred from the reconstructed power spectrum, could be used again to seed new mocks used in a new transfer function calculation, thus achieving enhanced accuracy. This is another item we leave for follow-up work, extending beyond the initial discussion and tests in \appref{sec:MCMCTF}.

\section{Conclusions}\label{sec:Conclusion}

\hi\ intensity mapping has the potential to be a leading resource for precision cosmology. A major challenge involves removing astrophysical foregrounds that dominate the underlying \hi\ cosmological signal by several orders of magnitude. Simulations and real observations are providing evidence that foregrounds can be sufficiently cleaned using blind separation techniques. However, quantifying the precise impact foreground cleaning has on the \hi\ power spectrum is crucial for avoiding biased analyses. In this work, we validate a method involving mock signal injection into the observed data as a means for accurately estimating the signal loss induced in the \hi\ as a consequence of the foreground clean. This method, referred to as the foreground transfer function, has been used in real data analysis before and its accuracy was validated for the results in \citet{Cunnington:2022uzo}, but its reliability for the purposes of precision cosmology has not been studied until now. For the first time, we present simulation-based tests demonstrating the foreground transfer function's accuracy as a tool for reconstructing estimated \hi\ power spectra to sub-percent accuracies.

This work \textcolor{black}{used} a selection of simulations to enable tests on a range of scenarios. In all cases we used an underlying \hi\ intensity map generated from an $N$-body simulation, from which we could measure the "true" signal. Simulated foreground maps were then added and a PCA-based foreground clean was performed, the consequences of which were analysed relative to the truth. We varied the complexity of the foregrounds and observational effects providing scenarios with differing demands from the foreground clean, hence presenting a range in signal loss. Foreground transfer functions were constructed using lognormal \hi\ intensity mapping mocks and their ability to recover the true signal could be scrutinised. In this work, our focus was on \textit{signal loss} from over-cleaning, as opposed to the other undesirable consequence of \textit{foreground residuals} caused by under-cleaning. Foreground residuals should be reducible to sub-dominant contributions or circumvented in cross-correlation with e.g. galaxy surveys, hence are less problematic than signal loss for precision cosmology. In the majority of tests, we therefore inspected the reconstructed cross-correlation of the cleaned intensity map with the original foreground-free (\hi-only) map. This way, only the effects from signal loss would manifest and the impact from foreground residuals would be mitigated, which cause a positive bias in the \hi\ auto-correlation. 

We summarise our main conclusions below;

\begin{itemize}[leftmargin=*]

\item We summarised the recipe for estimating a foreground transfer function (\secref{sec:TFsummary}) using mock signal injection into the observed data, which delivers an unbiased reconstruced power spectrum (\autoref{fig:TFdifferentcases}). These results included simulated polarisation leakage which demanded an aggressive foreground clean, resulting in ${>}\,50\%$ signal loss on the largest scales. In \autoref{fig:TFdifferentcasesBias} we demonstrated the potential consequences of deviating from this unbiased recipe resulting in under-estimated signal loss of up to ${\sim}\,30\%$. 
\\
\item We validated a technique for estimating the covariance in reconstructed power spectra which involves calculating the covariance across the reconstructed power spectra from all injected mock realisations (\autoref{fig:ErrorEst}). Crucially, when adopting this approach, the cleaned observed data $\textbf{\textsf{X}}_\text{clean}$ must not be subtracted when calculating the transfer function, which is otherwise removed to reduce the variance in the transfer function and optimise its accuracy.
\\
\item It has been previously assumed that the transfer function should be applied twice to correct for auto-correlation signal-loss i.e. $P_\text{rec}^\text{auto}\txteq{=}P_\text{clean}^\text{auto}\mathcal{T}^{-2}$. However, our simulations tests show that this is not the case (\autoref{fig:AutoVsCross}). At the fundamental 3D Fourier transform level, the average suppression in signal is the same in auto-correlation and cross-correlations with a foreground-free tracer (\autoref{fig:DiscreteSignalLoss}). Thus a $\mathcal{T}^{-1}$ transfer function is the correct reconstruction in both cross- and auto-correlation power spectra. 
\\
\item When calculating the transfer function, we estimated power directly in 1D spherically averaged $k$-bins, finding this performed well on all our simulations. This deviates from some previous approaches which instead calculate and apply a transfer function in 2D cylindrically averaged $k_\perp,k_\parallel$-bins, then re-projects these bandpowers into 1D $k$-bins. However, we found evidence that this made results prone to a higher variance (\autoref{fig:AnisotropicTF}), which may require a more detailed weighted average in the 2D to 1D projection to mitigate the issue.
\\
\item To "stress-test" the transfer function performance, we applied it on a pathfinder-like intensity map where the survey volume is relatively small, systematic effects are present, and signal-to-noise is low. To do this we produced the MDMK simulation that emulated the MeerKAT 2019 pilot survey using the same footprint, non-uniform frequency coverage, and perturbed foreground spectra, produced using the MeerKAT data itself (\secref{sec:MDMK}). Despite these increased challenges which result in more widespread signal loss, the transfer function was still able to reconstruct an unbiased power spectrum (\autoref{fig:MKpower}), validating the approach in \citet{Cunnington:2022uzo}.
\\
\item Finally, we confirmed how the transfer function accuracy has a relatively low dependency on the input parameters used for the mock generation in the transfer function calculation. To demonstrate this we chose some example parameters and biased them relative to their true fiducial values. Even going to extreme $+/-\,100\%$ biases, yielded small biases in the reconstructed power spectra and ${<}\,5\%$ bias in the recovered parameter estimates relative to the foreground-free case (\autoref{fig:ParameterResponse}). We also ran an extreme test where the underlying fiducial cosmology had an $f_\text{NL}\txteq{=}100$ value producing a scale-dependent bias on the largest scales. Despite a foreground clean heavily distorting this PNG signature, we demonstrated that an unbiased recovery is still obtained even if assuming $f_\text{NL}\txteq{=}0$ in the transfer function calculation (\autoref{fig:fNL}). 
\end{itemize}
These results place increased confidence in using a foreground transfer function as \hi\ intensity mapping ventures into precision cosmology. There is also some flexibility in regard to the dependency on reconstruction. Where levels of signal loss are high (e.g. ${>}\,50\%$ on large scales), post-cleaning scale cuts can be imposed to limit the dependency on the reconstruction. We discussed this in \secref{sec:TFsummary} and argued that excluding small-$k_\parallel$ modes from the analysis will massively reduce signal loss. A transfer function is still an essential tool in these scenarios since some reconstruction across all scales will be required. Importantly, the transfer function will also estimate the regions where signal loss is high, informing scale cut locations.

We have strived to demonstrate our results on a range of simulations with differing observational effects and survey sizes. However, simulations can have approximations not present in reality, thus validating the transfer function for signal loss reconstruction will remain an ongoing pursuit with more specific tests. One relevant extension we discussed will be to use simulations with a complex beam pattern, as opposed to the Gaussian beam assumed in this work. An incorrect beam model can have a high impact beyond the transfer function hence, we decided to leave this to a future more dedicated study, extending on the work of \citet{Matshawule:2020fjz,Spinelli:2021emp}. We will also aim to include $1/$f noise in the simulations which can impact foreground cleaning \citep{Harper:2017gln,Li:2020bcr,Irfan:2023njr}. Investigating how the transfer function can be constructed and applied in other clustering estimators will also be crucial. For example, extending the transfer function formalism to higher-order multipoles, applying it to correct the quadrupole and hexadecapole. There also needs to be investigation into whether it can be applied in configuration space to the correlation function or whether it can be used for correcting foreground cleaning effects in $n$-point statistics such as the bispectrum \citep{Cunnington:2021czb}.

There is continual improvement in foreground removal \citep{Makinen:2020gvh,Soares:2021ohm,Irfan:2021bci,Gao:2022xdb} mostly owing to machine learning and it may be that blind routines and the signal loss they cause become obsolete. Forward modelling frameworks have also been proposed as a means for reconstructing foreground contaminated modes \citep{Modi:2019hnu}. Furthermore, there is a possibility that future surveys and understanding will be sufficiently sophisticated to allow precise modelling of the foregrounds without requiring removal \citep{Fonseca:2020lmi}. However, the reliability of these methods on real data is yet to be showcased and it is likely that blind foreground removal techniques will be the preferred method for some time. It is therefore crucial that we continue to understand how to correct for the signal loss they cause.

\section*{Acknowledgements}

The authors would like to thank Jos\'{e} Luis Bernal, Zhaoting Chen, Aishrila Mazumder, Azadeh Moradinezhad Dizgah, Paula S. Soares and Amadeus Wild for helpful discussions. SC also would like to thank Isabelle Ye, Georgia Kiddier and Dounia Lacroze all of whom pursued masters projects testing the foreground transfer function, supplying plenty of thought-provoking results. \textcolor{black}{We also thank Matilde Barberi Squarotti for noticing a typo in the pre-print relating to the power spectrum model equation.}

SC is supported by a UK Research and Innovation Future Leaders Fellowship grant [MR/V026437/1]. LW  is a UK Research and Innovation Future Leaders Fellow [MR/V026437/1]. This result is part of a project that has received funding from the European Research Council (ERC) under the European Union's Horizon 2020 research and innovation programme (Grant agreement No. 948764; PB). PB acknowledges support from STFC Grant ST/T000341/1.
IPC\ acknowledges support from the `Departments of Excellence 2018-2022' Grant (L.\ 232/2016) awarded by the Italian Ministry of University and Research (\textsc{miur}) and from the `Ministero degli Affari Esteri della Cooperazione Internazionale (\textsc{maeci}) -- Direzione Generale per la Promozione del Sistema Paese Progetto di Grande Rilevanza ZA18GR02.
AP is a UK Research and Innovation Future Leaders Fellow
[grant MR/S016066/2].
MS acknowledges support from the AstroSignals Synergia grant CRSII5\_193826 from the Swiss National Science Foundation.
MGS and MI acknowledge support from the South African Radio Astronomy Observatory and National Research Foundation (Grant No. 84156)
We acknowledge the use of the Ilifu cloud computing facility, through the Inter-University Institute for Data Intensive Astronomy (IDIA). The MeerKAT telescope is operated by the South African Radio Astronomy Observatory, which is a facility of the National Research Foundation, an agency of the Department of Science and Innovation.

For the purpose of open access, the authors have applied a Creative Commons Attribution (CC BY) licence to any Author Accepted Manuscript version arising.
\section*{Data Availability}

The data underlying this article will be shared on reasonable request to the corresponding author.



\bibliographystyle{mnras}
\bibliography{Bib} 




\appendix

\section{Simulated Intensity Map Data}\label{sec:sims}

In this work, we use three different simulations to allow testing of the transfer function under different scenarios. These three simulations are referred to as
\begin{enumerate}[wide, label=\arabic*., labelwidth=!, labelindent=0pt]
    \item \textbf{MD1GPC}: The standard \textsc{Multi-Dark} $1\,\text{Gpc}^3h^{-3}$ simulation we used as the default in the majority of the paper unless otherwise mentioned.
    \\
    \item \textbf{MDMK}: \textsc{Multi-Dark} simulations but applied to a MeerKAT pilot-survey footprint to test applications of the transfer function in data representative of current single-dish intensity maps. Used for the investigation in \secref{sec:MK} and the details of its construction are explained there.
    \\
    \item \textbf{COLA}: $N$-body simulation which can be run for large physical dimensions to allow investigation into the robustness of transfer function on ultra-large-scales in future data sets. Used only for the $f_\text{NL}$ investigation in \secref{sec:fNL}.
\end{enumerate}

\noindent The MDMK and COLA simulations are mostly extensions of the MD1GPC simulations, explained in the relevant sections (\secref{sec:MK} and \secref{sec:fNL} respectively). The details of the default MDMK simulation are outlined below.

\subsection{\textsc{Multi-Dark} \hi\ simulation (MD1GPC)}\label{sec:MD1GPC}

For our main simulated \hi\ intensity maps, which are used in all parts of the paper unless clearly stated, we use the same simulations as those adopted in \citep{Cunnington:2020njn,Cunnington:2021czb}. These used the \textsc{MultiDark-Galaxies} $N$-body simulation data \citep{Knebe:2017eei} and the catalogue produced from the \textsc{SAGE} \citep{Croton:2016etl} semi-analytical model application. These galaxies were produced from the dark matter cosmological simulation \textsc{MultiDark-Planck} (MDPL2) \citep{Klypin:2014kpa}, which follows the evolution of 3840$^3$ particles in a cubical volume of $1\,(\text{Gpc}/h)^3$ with mass resolution of $1.51\times10^9h^{-1}$M$_\odot$ per dark matter particle. The cosmology adopted for this simulation is based on \textsc{Planck}15 cosmological parameters \citep{Ade:2015xua}, with $\Omega_\text{m}\txteq{=}0.307$, $\Omega_\text{b}\txteq{=}0.048$, $\Omega_\Lambda\txteq{=}0.693$, $\sigma_8\txteq{=}0.823$, $n_\text{s}\txteq{=}0.96$ and Hubble parameter $h=0.678$. The catalogues
are split into 126 snapshots between redshifts $z\txteq{=}17$ and $z\txteq{=}0$. In this work we chose low-redshift, post-reionisation data to test the transfer function and use the snapshot at $z\txteq{=}0.39$ to emulate a MeerKAT-like survey performed in the L-band ($899 \txteq{<}\nu\txteq{<}1184\,\text{MHz}$, or equivalently $0.2\txteq{<}z\txteq{<}0.58$). Although there is no reason to suspect conclusions will be any different for any reasonable redshift choice between $0\txteq{<}z\txteq{<}3$. We obtained this publicly available data from the Skies \& Universes web page\footnote{\href{http://www.skiesanduniverses.org/page/page-3/page-22/}{www.skiesanduniverses.org}}. 

We used each galaxies (x, y and z) coordinates and placed them onto a grid with $n_\text{x}, n_\text{y}, n_\text{z}\txteq{=}256,256,256$ pixels and $1\,(\text{Gpc}/h)^3$ in physical size. To simulate observations in redshift space inclusive of RSD, we \textcolor{black}{used} the peculiar velocities of the galaxies. Assuming the LoS is along the z-dimension and given the plane-parallel approximation is exact for this Cartesian data, RSD can be simulated by displacing each galaxy's position to a new coordinate $z_\text{RSD}$ given by $\text{z}_\text{RSD}\txteq{=}\text{z}\txteq{+}v_\parallel(1\txteq{+}z)h/H(z)$, where $v_\parallel$ is the galaxy's peculiar velocity along the LoS (z-dimension) which is given as an output of the simulation in units of $\text{km}\,\text{s}^{-1}$.

To simulate the contribution to the signal from each galaxy, we used the cold gas mass $M_\text{cgm}$ output from the \textsc{MultiDark} data and from this we can infer a \hi\ mass with $M_\hi \txteq{=}f_\text{H} M_\text{cgm}(1-f_\text{mol})$ where $f_\text{H}=0.75$ represents the fraction of hydrogen present in the cold gas mass and the molecular fraction is given by $f_\text{mol}\txteq{=}R_\text{mol}/(R_\text{mol}+1)$ \citep{Blitz:2006nc}, with $R_\text{mol}\txteq{\equiv}M_{H_2}/M_\hi\txteq{=}0.4$ \citep{Zoldan17}. It is this \hi\ mass that we binned into each voxel with position $\boldsymbol{x}$, to generate a data cube of \hi\ masses $M_\hi(\boldsymbol{x})$,  which should trace the underlying matter density generated by the catalogue's $N$-body simulation for the snapshot redshift $z$. These \hi\ masses are converted into a \hi\ brightness  temperature for a frequency width of $\deltadiff \nu$ subtending a solid angle $\deltadiff \Omega$ given by
\begin{equation}\label{THIequation}
    T_\hi(\boldsymbol{x},z) = \frac{3h_\text{P}c^2A_{12}}{32\pi m_\text{h}k_\text{B}\nu_{21}}\frac{1}{\left[(1+z)r(z)\right]^2}\frac{M_\hi(\boldsymbol{x})}{\deltadiff \nu \, \deltadiff \Omega} \, ,
\end{equation}
where $h_\text{P}$ is the Planck constant, $A_{12}$ the Einstein coefficient that quantifies the rate of spontaneous photon emission by the hydrogen atom, $m_\text{h}$ is the mass of the hydrogen atom, $k_\text{B}$ is Boltzmann's constant, $\nu_{21}$ the rest frequency of the 21cm emission and $r(z)$ is the comoving distance out to redshift $z$ (we will assume a flat universe). Since \hi\ simulations on this scale have a finite halo-mass resolution, there will be some contribution from the \hi\ within the lowest-mass host haloes which is not included in the final $T_\hi$ signal. To account for this, it is typical for a rescaling of the final $T_\hi$ to be performed to bring the mean \hi\ temperature, $\overline{T}_\hi$, in agreement with the modest data constraints we have for this value. For the effective redshift of our data, $z\txteq{=}0.39$, we used a fiducial value of $\overline{T}_\hi\txteq{=}0.0743\,\text{mK}$ which our maps were re-scaled to.

\subsection*{MD1GPC Foreground Simulations}\label{app:FGSims}

To produce the foreground contamination we simulated different foreground processes, including galactic synchrotron, free-free emission and point source emission. We also included the effects of polarisation leakage which will act as an extra component of foreground with non-smooth spectra \citep{Cunnington:2020njn}, thus posing an increased challenge for the foreground clean. The foregrounds we used can thus be decomposed as $T_\text{fg}\txteq{=}T_\text{sync} + T_\text{free} + T_\text{point} + T_\text{pol}$, which represent the synchrotron, free-free, point sources and polarisation leakage.

We briefly summarise the simulation technique for these components but for a full outline we refer the reader to \citet{Cunnington:2020njn} and \citet{Carucci:2020enz} where they were also used. The synchrotron emission is based on Planck Legacy Archive\footnote{\href{http://pla.esac.esa.int/pla}{pla.esac.esa.int/pla}} FFP10 simulations of synchrotron emission at $217$ and $353\,\text{GHz}$ formed from the source-subtracted and destriped $0.408\,\text{GHz}$ map. The free-free simulation is from the FFP10 217\,GHz free-free simulation which is a composite of the \citet{cliveff} free-free template and the WMAP MEM free-free templates. The point sources are based on the empirical model of \citet{batps} and makes the assumption that point sources over $10\,\text{mJy}$ will be identifiable and thus can be removed. Lastly, we simulated polarisation leakage with the use of the \texttt{CRIME}\footnote{\href{http://intensitymapping.physics.ox.ac.uk/CRIME.html}{intensitymapping.physics.ox.ac.uk/CRIME.html}} software \citep{Alonso:2014sna}, which provides maps of Stokes Q emission at each frequency and we fix the polarisation leakage to $0.5\%$ of the Stokes Q signal.

For the foregrounds we assumed they have been observed in a frequency range of $900\txteq{<}\nu\txteq{<}1156\,\text{MHz}$, consistent with the $z\txteq{=}0.39$ redshift for the cosmological simulation. Each of the 256 map slices along the z-direction acts as an observation in a frequency channel giving a channel width of $\deltadiff \nu \txteq{=} 1\,\text{MHz}$. This therefore emulates the spectral distinction between the cosmological \hi\ and foregrounds utilised in the foreground clean. From the full-sky foreground map we cut a region of sky centred on the Stripe~82, a field well observed by surveys. The size of this sky region is $54.1\txteq{\times}54.1\,\text{deg}^2$ which corresponds to the size of a $1\,(\text{Gpc}/h)^2$ patch at the $z\txteq{=}0.39$ snapshot redshift of our cosmological simulation.

\subsection{Instrumental effects}

Here we outline some of the additional observational effects which we switch on and off for different scenarios throughout the paper. Unless clearly mentioned, the reader can assume these effects have not been included for simplicity.

\subsection*{Telecope beam}

The effect from the telescope beam is a smoothing to the temperature field in directions perpendicular to the LoS. A simple, and often sufficient, method to simulate these beam effects is to convolve the density field with a Gaussian kernel whose FWHM ($\theta_\text{FWHM}$) is chosen to match the model of the radio telescope one is trying to emulate. We can define this Gaussian smoothing kernel with
\begin{equation}\label{eq:GaussianSpatialBeam}
\begin{aligned}    
    \mathcal{B}_\text{G}(\nu,\boldsymbol{s}_{\perp}) &= \exp\left[-4\ln2 \left(\frac{\boldsymbol{s}_{\perp}}{r(\nu)\,\theta_\text{FWHM}(\nu)} \right)^2 \right]\\
    &= \exp\left[\frac{1}{2} \left(\frac{\boldsymbol{s}_{\perp}}{R_\text{beam}} \right)^2 \right]\,,
\end{aligned}
\end{equation}
where $\boldsymbol{s}_{\perp} \txteq{=} \sqrt{\Delta x^2 + \Delta y^2}$ is the perpendicular spatial separation from the centre of the beam. $R_\text{beam}\txteq{=}r(z)\,\sigma_\text{beam}$ defines the physical size of the beam's central lobe in Mpc/$h$, where $\sigma_\text{beam}\txteq{=}\theta_{\mathrm{FWHM}} /(2 \sqrt{2 \ln 2})$ represents the standard deviation of the Gaussian kernel in radians. $R_\text{beam}$ is dependent on frequency through the comoving distance out to the density fluctuations which changes with frequency ($r(\nu)$). It also has a further frequency dependence from the intrinsic beam size of the instrument, which is itself a function of frequency, generically given by $\theta_{\mathrm{FWHM}} \txteq{\approx} c/v D_{\mathrm{dish}}$, where $D_{\mathrm{dish}}$ is the diameter of the radio telescope dish.

As we discussed in the main body of the paper, including a more sophisticated model of the beam is worth investigating since this could have implications for foreground removal, signal loss, and thus signal reconstruction. A more complex beam with far-reaching side lobes and frequency dependence has been shown to create additional challenges for foreground cleaning \citep{Matshawule:2020fjz,Spinelli:2021emp}. This is an upgrade which will be targeted in future work.

\subsection*{Instrumental Noise}

An unavoidable source of noise in intensity mapping comes from the thermal motion of electrons inside the electronics of the instrument which produce Gaussian-like fluctuating currents, with a mean current of zero but a non-zero rms. The consequence from this is a component of white-noise contained in the maps. From the radiometer equation, the rms of the thermal noise contained in time-ordered data for an instrument with system temperature $T_\text{sys}$, frequency resolution $\delta\nu$ and time per pointing $t_\text{p}$, will be given by \citep{Wilson2009}
\begin{equation}\label{eq:noise}
    \sigma_\text{n} = T_\text{sys} \Big/ \sqrt{2\,\delta\nu\,t_\text{p}}\,.
\end{equation} 
At map level this will create a field of white noise added into the data, with rms $\sigma_\text{n}$. In the case of the power spectrum this produces an additive component;
$P_\hi \rightarrow P_\hi + P_\text{N}$ where
$P_\text{N} = \sigma_\text{n}^2/V_\text{cell}$. Since this should be uncorrelated and independent at different observation times and for different dishes, this thermal noise can be averaged down as survey time increases. Thus, it is not seen as a major problem for future intensity mapping surveys where long observation campaigns will be conducted. In all cases where we include noise in the simulations we use Gaussian white noise with a value of $\sigma_\text{n}\txteq{=}1\,\text{mK}$. This is designed to dominate over the \hi\ which has an rms of $\sigma_\hi\txteq{\sim}0.14\,\text{mK}$. The time per pointing is defined as
\begin{equation}\label{eq:timeperpoint}
    t_{\mathrm{p}}=N_{\mathrm{dish}} t_{\mathrm{obs}}\left(\theta_{\mathrm{FWHM}} / 3\right)^2 / A_{\mathrm{sky}}\,,
\end{equation}
where we have assumed the pixel size will be 1/3 of the beam size. For a MeerKAT-like $A_\text{sky}\txteq{\sim}3{,}000\,\text{deg}^2$ survey with $N_\text{dish}\txteq{=}64$ dishes, $\deltadiff\nu\txteq{=}0.2\,\text{MHz}$ frequency resolution and $T_\text{sys}\txteq{=}16\,\text{K}$ \citep{Wang:2020lkn}, the $\sigma_\text{n}\txteq{=}1\,\text{mK}$ dominant noise will correspond to $t_\text{obs}\txteq{\sim}30\,\text{hrs}$ of observation time. 

\section{Power spectrum estimation}\label{sec:PowerSpec}

Here we briefly outline the method for measuring power spectra, used throughout the paper. This follows the same methodology as the MeerKAT intensity mapping pipeline \citep{Cunnington:2022uzo}.

We define the Fourier transform of the \hi\ intensity maps $\delta T_\hi$ as
\begin{equation}
    \tilde{F}_\hi(\boldsymbol{k})=\sum_{\boldsymbol{x}} \delta T_\hi(\boldsymbol{x}) w_\hi(\boldsymbol{x}) \exp (i \boldsymbol{k}{\cdot}\boldsymbol{x})\,,
\end{equation}
where $w_\hi$ are the weights that can be applied to optimise the power spectrum measurement. For our simulations we simply assume $w_\hi\txteq{=}1$ everywhere. The \hi\ power spectrum is then estimated by
\begin{equation}\label{eq:HIautoPk}
    \hat{P}_\hi(\boldsymbol{k}) = \frac{V_\text{cell}}{\sum\limits_{\boldsymbol{x}} w_\hi^2(\boldsymbol{x})}\lvert\tilde{F}_\hi(\boldsymbol{k})\rvert^2\,.
\end{equation}
In many of the results, we use the cross-correlations between a \hi-only (foreground-free) simulation $\tilde{F}_\hi$ and a foreground-cleaned one $\tilde{F}_\text{clean}$. In this case the cross-correlation is similarly defined by
\begin{equation}\label{eq:HIcrossPk}
    \hat{P}_\text{X}(\boldsymbol{k}) = \frac{V_\text{cell}}{\sum\limits_{\boldsymbol{x}} w^2_\hi(\boldsymbol{x})}\operatorname{Re}\left\{\tilde{F}_\hi(\boldsymbol{k}){\cdot} \tilde{F}^{*}_\text{clean}(\boldsymbol{k})\right\}\,.
\end{equation}
These power spectra are either spherically averaged into bandpowers $\lvert\boldsymbol{k}\rvert\,{\equiv}\,k$ to provide the 1D power spectra results, or cylindrically averaged into $k_\perp,k_\parallel$ bins to produce the demonstrative 2D power spectra plots.

\subsection{Modelling the \hi\ intensity mapping power spectrum}\label{sec:HIPkMod}

Wherever we require a model for the observational simulations, we use the below prescription;
\begin{multline}\label{eq:Pkmodel}
    P_\text{mod}(k,\mu) = \overline{T}^2_\hi \frac{\left(b^2_\hi+f\mu^2\right)^2}{1+(k\mu\sigma_\text{v}/H_0)^2} P_\text{m}(k)\\
    \times\,\exp\left[-(1-\mu^2)k^2 R_\text{beam}^2\right]\,,
\end{multline}
where $b_\hi$ is the linear bias for the \hi\ field and $\overline{T}_\hi$ is the mean \hi\ temperature in mK and approximately related to the \hi\ density fraction by
\begin{equation}\label{eq:TbarModelEq}
    \overline{T}_\hi(z) = 180\,\Omega_{\hi}(z)\,h\,\frac{(1+z)^2}{\sqrt{\Omega_\text{m}(1+z)^3 + \Omega_\Lambda}} \, {\text{mK}} \,,
\end{equation}
where $\Omega_\text{m}$ and $\Omega_\Lambda$ are the density fractions for matter and the cosmological constant, respectively. The linear redshift-space distortions (RSD) are accounted for in \autoref{eq:Pkmodel} by the $f\mu^2$ term \citep{Kaiser:1987qv}, where $f$ is the growth rate of structure and $\mu$ is the cosine of the angle from the line-of-sight. In the denominator, we approximately account for the non-linear effects of RSD, commonly referred to as Fingers-of-God, and $\sigma_\text{v}$ is the velocity dispersion of the \hi, with $H_0$ as the Hubble constant. $P_\text{m}$ is the matter power spectrum produced using \texttt{CAMB}\footnote{\href{https://camb.readthedocs.io/en/latest/}{camb.readthedocs.io/en/latest/}} \citep{Lewis:1999bs} with a Planck18 \citep{Aghanim:2018eyx} cosmology. The exponential factor accounts for the smoothing of perpendicular modes due to the beam, where $R_\text{beam}$ is the standard deviation of the Gaussian beam profile in comoving units, as explained in \appref{sec:sims}.

\section{Examples of Biased transfer functions}\label{sec:BiasedTF}

\begin{figure}
    \centering
    \includegraphics[width=1\linewidth]{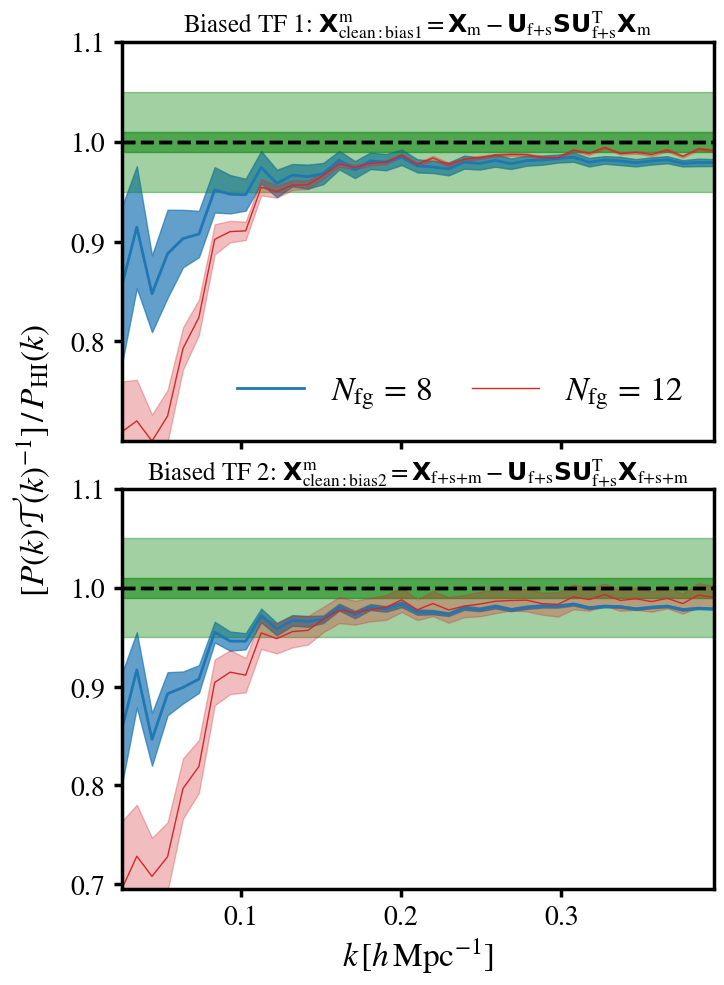}
    \caption{Same as \autoref{fig:TFdifferentcases} but for versions of the transfer function that deliver biased results. The transfer functions versions vary based on their definition of $\textbf{\textsf{X}}^\text{m}_\text{clean}$, which are defined by the panel titles in each version and are discussed further in the appendix text. All transfer functions are calculated by averaging over 100 lognormal mocks and the shaded bands show the rms over these 100 mocks. For each version, results for a mild ($\Nfg\txteq{=}8$, blue lines) and more aggressive ($\Nfg\txteq{=}12$, red lines) foreground cleans are shown. Dark (light) green regions indicate sub 1\% (5\%) accuracy of the reconstructed power spectrum.}
    \label{fig:TFdifferentcasesBias}
\end{figure}

As demonstrated in \secref{sec:signallossformalism} and explicitly highlighted in \citet{Switzer:2015ria}, when estimating the impact from blind foreground cleaning on the signal, it is insufficient to consider only the direct loss to the signal in the removed $\textbf{\textsf{U}}_\text{f}\textbf{\textsf{S}}\textbf{\textsf{U}}^\text{T}_\text{f}\textbf{\textsf{X}}_{\mathrm{s}}$ piece. One must also include the impact of spurious correlations caused by the presence of non-foreground data such as the \hi\ signal itself and the perturbations $\mathbf{\Delta}$ these cause to the estimated eigenmodes. This is particularly important when considering how to estimate signal loss using mock data. For example, adopting an approach which simply looks to project out the observed data modes $\textbf{\textsf{U}}_\text{f+s}$ from realisations of mock data $\textbf{\textsf{X}}_\text{m}$, will neglect the signal loss from the perturbed terms. Even though the foreground modes are perturbed by the true signal $\textbf{\textsf{X}}_\text{s}$, cross-terms from these perturbations will be uncorrelated with the mock signal $\textbf{\textsf{X}}_\text{m}$, which is what matters in the construction of the transfer function where signal loss to the mocks is being evaluated. As an example, if we assume the cleaned mocks can be defined by 
\begin{equation}\label{eq:Mcleanbias1}
    \textbf{\textsf{X}}^\text{m}_\text{clean:bias1} =\textbf{\textsf{X}}_\text{m} -  \textbf{\textsf{U}}_\text{f+s}\textbf{\textsf{S}}\textbf{\textsf{U}}_\text{f+s}^\text{T}\textbf{\textsf{X}}_\text{m}\,,
\end{equation}
then use this in the transfer function (\autoref{eq:TransferFunc}),
the resulting reconstructed power spectrum will be slightly biased (${>}\,1\%$) on small scales (as shown by the top panel of \autoref{fig:TFdifferentcasesBias}), and more extremely biased on larger scales, reaching $10\%$ departure from the truth at $k\txteq{\sim}0.1\hMpc$ for the $\Nfg\txteq{=}12$ case. This is because the source of the perturbations to the eigenmodes, which in this case is only the true signal $\textbf{\textsf{U}}_\text{f+s}\txteq{=}\textbf{\textsf{U}}_\text{f}\txteq{+}\mathbf{\Delta}_\text{s}$, is being projected out of data ($\textbf{\textsf{X}}_\text{m}$) which will have no correlation with these perturbations, thus this contribution is neglected, hence the bias. This also explains why the bias is worse for higher $\Nfg$, because the neglected correlations are larger for higher $\Nfg$ (shown by \autoref{fig:DecomposedTerms}).

A slight improvement can be attempted on \autoref{eq:Mcleanbias1} by projecting out the data modes $\textbf{\textsf{U}}_\text{f+s}$ over the true data with mock signal injected $\textbf{\textsf{X}}_\text{f+s+m}$
\begin{equation}\label{eq:Mcleanbias2}
    \textbf{\textsf{X}}^\text{m}_\text{clean:bias2} = \textbf{\textsf{X}}_\text{f+s+m} - \textbf{\textsf{U}}_\text{f+s}\textbf{\textsf{S}}\textbf{\textsf{U}}_\text{f+s}^\text{T}\textbf{\textsf{X}}_\text{f+s+m}\,. 
\end{equation}
However, as the bottom panel of \autoref{fig:TFdifferentcasesBias} shows, the bias is still present since it still lacks any correlation in perturbed modes and mock signal, thus fails to emulate the correlations between signal and foregrounds.

These results from the biased versions of the transfer function thus highlight the importance of emulating the spurious correlations between foregrounds and signal in the construction of the transfer function. The correct approach from \autoref{eq:Mclean} should thus always be adopted. Along with the discussion of this point in \citet{Switzer:2015ria}, it has also been investigated in epoch of reionisation studies \citep{Cheng:2018osq} where it was acknowledged how neglecting these additional complications leads to an under-estimation of the signal loss. 

\section{Iterative foreground transfer function \& a Bayesian approach?}\label{sec:MCMCTF}

Our investigation which varied the input parameters for the mocks used in the transfer function construction provided encouraging results (demonstrated by \autoref{fig:ParameterResponse} and \autoref{fig:fNL}). This showed that the transfer function only has a very mild dependence on the mock input parameters. For the parameters we tested, we found they can be highly biased relative to the truth in observed data, but this does not have a significant impact on the reconstructed power spectra. However, some dependence was still noticed, and if one is striving to maximise accuracy in parameter inference then this dependence may be enough to cause concern. 

We raised the idea in the main text of an \textit{iterative} transfer function. Since the reconstructed power spectra show good agreement with the truth despite large mock parameter biases, there is a strong possibility that the parameters inferred from a reconstructed power spectrum will be much closer to the truth. By using these updated parameter estimates to seed new mocks and construct a new transfer function, it is highly likely the new reconstructed power spectrum based on the updated transfer function will have an even better agreement with the truth. This process could then be repeated indefinitely, iteratively improving the accuracy of the transfer function until convergence on all inferred parameters is achieved. 

\begin{figure}
    \centering
    \includegraphics[width=1\linewidth]{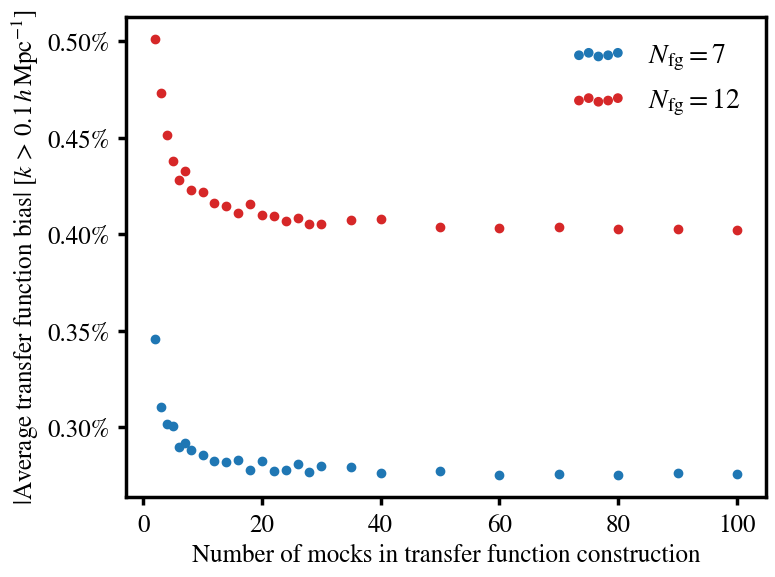}
    \caption{Number of mocks required in transfer function computation to reach a converged level of accuracy in the reconstructed power spectrum for two levels of foreground cleaning given by $\Nfg$. The $y$-axis shows the mean bias for each number of mocks in the reconstructed power spectrum at $k\txteq{>}0.1\hMpc$ values relative to the truth (\hi-only power). We average the \textit{absolute values} of the the power spectrum biases to avoid potential cancellation to zero in highly fluctuating results around zero. For each number of mocks tested, we average over 100 realised combinations to get a stable accuracy level.}
    \label{fig:NumberofMocksConvergence}
\end{figure}

A challenge for an iterative transfer function is the computational demand required to repeatedly calculate one. The larger surveys become, the larger the mocks need to be thus computational expense will only increase. Thus, time wasted on mock generation when convergence has already been reached would be an unnecessary bottleneck. To provide some guidance on the issue we consider how many mocks are required for a stable transfer function. In \autoref{fig:NumberofMocksConvergence} we investigate how many mocks are required before the accuracy of the transfer function reaches a converged level. We define the accuracy of the transfer function by the average bias across $k$-modes for the reconstructed power spectrum relative to the original \hi-only power i.e. $[P(k)\mathcal{T}(k)^{-1}]/P_\hi(k)$. In this accuracy calculation, we ignore large scales ($k\txteq{<}0.1\hMpc$) where accuracy fluctuations can be quite large. We calculate the transfer function using different numbers of mocks and for each number tested we average over 100 iterations so that the returned accuracy is stable. 

\autoref{fig:NumberofMocksConvergence} suggests that when using just two mocks to construct the transfer function, excellent accuracy is already achievable, although these results would be prone to fluctuating performance based on the two mocks used each time. We see that accuracy can be improved by increasing the number of mocks but convergence is quickly reached at around 20. For the different levels of foreground clean, shown by the different $\Nfg$ results, the accuracy levels differ with the higher $\Nfg$ returning poorer accuracy as expected due to the increased contribution from the spurious foreground and signal correlations for higher $\Nfg$. However, we see that convergence is reached at a consistent number of mocks for both $\Nfg$ cases. Despite the evidence suggesting that only 20 mocks would be needed for any reasonably foreground clean, this would need to be tested further in real cases where the data could be more contaminated and thus may require a higher number of mocks for convergence. However, even in reality, the convergence level could be tested to ensure a Bayesian analysis is not computing an unnecessary number of mocks. What is encouraging from \autoref{fig:NumberofMocksConvergence} is how quick convergence is reached even for our fairly complex foreground simulations. We remind the reader that these mocks are only lognormal mocks and thus can be generated rapidly.


\bsp	
\label{lastpage}
\end{document}